%% file: IEEE-Jornal.tex
\documentclass {IEEEtran}
\IEEEoverridecommandlockouts
\usepackage{setspace}
\usepackage{soul}      
\usepackage{xcolor}    

\usepackage[mathlines,switch]{lineno}
\IEEEoverridecommandlockouts
\usepackage[mathlines,switch]{lineno}
\usepackage{etoolbox}
\AtBeginEnvironment{equation}{\linenomath}
\AtEndEnvironment{equation}{\endlinenomath}
\AtBeginEnvironment{align}{\linenomath}
\AtEndEnvironment{align}{\endlinenomath}
\AtBeginEnvironment{gather}{\linenomath}
\AtEndEnvironment{gather}{\endlinenomath}
\AtBeginEnvironment{IEEEeqnarray}{\linenomath}
\AtEndEnvironment{IEEEeqnarray}{\endlinenomath}
\AtBeginEnvironment{IEEEeqnarray*}{\linenomath}
\AtEndEnvironment{IEEEeqnarray*}{\endlinenomath}

\usepackage{soul}
\soulregister{\caption}{1}
\usepackage{cite}
\usepackage{graphicx}
\usepackage{amssymb, amsmath}
\usepackage[linesnumbered,ruled,vlined]{algorithm2e}
\usepackage{bm} %
\usepackage[switch]{lineno}
\usepackage{hyperref}
\usepackage{float}
\usepackage{booktabs}
\usepackage{subcaption}
\usepackage[T1]{fontenc}
\usepackage{newtxtext,newtxmath}
\usepackage{graphicx,calc,xcolor,amssymb,amsmath,mathrsfs,dsfont,hyperref,epstopdf}
\usepackage{xcolor}
\usepackage{caption}

\title{Sustainable Vertical Heterogeneous
Networks: A Cell Switching Approach with High Altitude Platform Station}

\begin{document}

 \author{\rm{Maryam Salamatmoghadasi}, \rm{Amir Mehrabian}, \rm{Halim Yanikomeroglu}, and \rm{Georges Kaddoum}
 \thanks{
Maryam Salamatmoghadasi and Halim Yanikomeroglu are with the Non-Terrestrial Networks Laboratory, Department of Systems and Computer Engineering, Carleton University, Ottawa, ON K1S5B6, Canada.
 (e-mails: \texttt{maryamsalamatmoghad@cmail.carleton.ca,  halim@sce.carleton.ca). }
 \\
 Amir Mehrabian and Georges Kaddoum are with the LaCIME Laboratory,
Department of Electrical Engineering, École de Technologie Supérieure,
Montreal, QC H3C 0J9, Canada. (e-mails: \texttt{\{amir.mehrabian, georges.kaddoum\}@etsmtl.ca).}
}
}

\maketitle
\begin{abstract}
The rapid growth of radio access networks (RANs), driven by evolving wireless technologies and increasing mobile traffic, poses significant energy consumption challenges and threatens the sustainability of future networks. In this context, vertical heterogeneous networks (vHetNets) offer a promising architectural solution to address these challenges. The present study explores a vHetNet configuration featuring a high altitude platform station (HAPS) that acts as a super macro base station (SMBS), along with a macro base station (MBS) and multiple small base stations (SBSs), in dense urban environments.
We propose a novel HAPS-enhanced cell switching (CS) algorithm that optimizes network performance by selectively deactivating SBSs, considering their traffic load as well as the capacity and channel conditions of both the MBS and HAPS. The algorithm addresses a mixed-integer nonlinear programming (MINLP) problem, reformulated into a mixed-integer programming (MIP) problem, to reduce vHetNet energy consumption while preserving an outage-based quality of service (QoS) constraint.
The novelty of this work lies in the formulation of a HAPS-enhanced CS framework and its evaluation under realistic 3GPP channel models. This combination differentiates our study from previous studies that either focused solely on terrestrial CS or relied on simplified propagation assumptions.
Extensive simulation results demonstrate that the HAPS-enhanced CS approach achieves substantial energy savings as compared to benchmark methods, including All-ON, Terrestrial CS, and Sorting. In particular, compared to All-ON, the proposed approach reduces power consumption by up to $77\%$ at low traffic loads and about $40\%$ at high loads. Moreover, the HAPS-enhanced CS NoQoS variant, which relaxes the outage-based QoS constraints to allow further SBS deactivation, achieves even larger reductions of up to $90\%$ at low load and $47\%$ at high load.
The algorithm also maintains high levels of served traffic, showing adaptability under varying load conditions and outage-based QoS requirements. By setting appropriate thresholds, the proposed algorithm effectively balances power efficiency and outage-based QoS, making it a scalable solution for sustainable 6G networks.

\end{abstract}
\begin{IEEEkeywords}
  vHetNet, HAPS, cell switching, sustainability, energy consumption
\end{IEEEkeywords}
\section{Introduction}
\subsection{Background}
Mobile network generations have evolved from the analog, voice-based first generation~(1G) networks to the vertical industry integrations of the fifth generation~(5G), which introduced features such as enhanced mobile broadband~(eMBB) and ultra-reliable low-latency communication~(URLLC). Each generation of networks has introduced fundamental new technologies, as well as improvements in key performance indicators like data rates, latency, and energy efficiency.

The sixth generation~(6G) is expected to build on previous advancements while introducing further transformative capabilities, particularly through the integration of non-terrestrial networks~(NTN). These technologies will address several limitations of terrestrial networks (TN), particularly in terms of coverage and operational flexibility, and offer enhanced energy efficiency and more effective resource management. A key element of NTN is a high altitude platform station (HAPS), which operates in the stratosphere at an altitude of approximately 20 km. HAPS provide stable, wide-area connectivity, which can significantly extend network coverage in remote or underserved regions, such as rural areas and disaster zones.

The International Telecommunication
Union (ITU) defines the HAPS as high-altitude International Mobile Telecommunications (IMT) base stations (HIBS)~\cite{itu_vision_november}. These stations have gained increased relevance after the identification of additional frequency bands for their operation, as confirmed by the recent World Radio Communication Conference 2023 (WRC-23)~\cite{WRC_23}. This enhancement allows HAPS to offer mobile broadband connectivity with minimal reliance on terrestrial infrastructure. By complementing macro base stations (MBSs), HAPS can play a critical role in addressing the capacity and energy efficiency challenges faced by 6G networks~\cite{maryam-mag, 11000303}. Recently, the World Economic Forum has recognized HAPS as one of the top 10 emerging technologies, underscoring their potential to revolutionize global connectivity and contribute to technological advancements~\cite{WEF}. These developments make HAPS an effective solution for improving coverage and ensuring more efficient, sustainable network infrastructure. By facilitating the deployment of low-cost, wide-area connectivity, HAPS are expected to significantly contribute to both 5G infrastructure and future 6G networks.

Along with improving network coverage, HAPS play a key role in achieving sustainability goals of future networks, particularly within the 6G vision outlined by the ITU in June 2023~\cite{itu_vision_june_23}. According to the European
Commission’s Industry 5.0 vision~\cite{industry5}, sustainability has become a fundamental design principle for 6G, which places HAPS at the core of future industrial and communication networks. Furthermore, recent resolutions from the World Radio Communication Conference 2023 (WRC-23)~\cite{WRC_23} place a particular emphasis on exploring innovative uses of the radio spectrum and promoting terrestrial-space convergence to achieve sustainable digital transformation.

The use of NTNs, and specifically HAPS, can be central to achieving these sustainability goals in 6G, as it allows for more energy-efficient and cost-effective operations by extending network coverage without the need for extensive terrestrial infrastructure. Integrating HAPS with existing TN is expected to improve energy efficiency, enhance global connectivity, and reduce the carbon footprint of cellular networks, which is consistent with the overarching goals of both 6G and Industry 5.0.

Energy efficiency is a critical focus area for future networks, driven by the increasing environmental impact of information and communication technologies (ICT). While, in 2020, the ICT industry accounted for approximately 4\% of global greenhouse gas emissions, this figure is projected to exceed 14\% by 2040~\cite{ICT}. As 6G is expected to support an unprecedented density of devices, especially in densely populated urban areas, the challenge of managing energy consumption becomes more acute, particularly in radio access networks~(RANs). At present, base stations (BSs) account for 60\% to 80\% of the total energy consumption in cellular networks~\cite{9340607,bs_power}. One promising solution to mitigate this energy usage is the implementation of \textit{cell switching (CS)}, which strategically deactivates or puts lightly used or idle BSs into sleep mode during periods of low traffic to conserve energy. This approach can substantially reduce network power consumption and there is growing evidence in the literature for its potential in 6G networks~\cite{bs_power, cell_switch_survey,8372969}.

\subsection{Related Work}
The CS mechanism has been widely explored in the literature~\cite{9528008, 8735834, ACM, 8024181,10061602, ELAA2022JR, EOMK2017JR, Metin_VFA_CellSwitch}, with a particular focus on reducing power consumption in cellular networks. 
In~\cite{9528008}, the authors investigated a BS switching-off strategy using traffic prediction for heterogeneous BSs with varying environmental characteristics in heterogeneous networks (HetNets), to minimize overall network power consumption. 
A BS switching-off algorithm was also proposed in~\cite{8735834}, where the authors aimed to reduce energy consumption by selectively activating a subset of BSs.
In~\cite{ACM}, four switching-off strategies were designed to save the energy consumption of the network. 
In~\cite{8024181}, the authors introduced a switching-off strategy, incorporating power adjustment to optimize energy efficiency in HetNets. Using this approach, BSs in small cells were selected to be switched off, while the MBS remained active to connect subscribers and provide service when necessary. 
Building on these studies, in~\cite{10061602}, the authors introduced an energy-efficient beamforming-aware BS switching-off method for multi-input single-output (MISO) cellular networks, optimizing energy efficiency through cooperative beamforming and roaming-cost-based sharing schemes.

A tiered sleep mode system that adjusts sleep depth based on device activity was proposed in~\cite{ELAA2022JR}, using decentralized control for scalability.
Furthermore, in~\cite{EOMK2017JR}, the authors explored the control data separated architecture (CDSA) and implemented a genetic algorithm to optimize energy savings in HetNets by managing user associations and BS deactivation through deterministic algorithms.
In another relevant study~\cite{Metin_VFA_CellSwitch}, the authors introduced a value function approximation (VFA)-based reinforcement learning (RL) algorithm for CS in ultra-dense networks, demonstrating scalable energy savings.

Available research on HAPS-supported CS is still in its early stages, with limited studies exploring this emerging area in depth.
Among the few relevant studies,~\cite{berk} proposed CS for HAPS-integrated TN using the exhaustive search (ES) algorithm.
Similarly, in~\cite{maryam}, the authors examined a vHetNet model with a HAPS-SMBS, an MBS, and several SBSs, focusing on optimizing the sleep mode management of SBSs by sorting their payloads in ascending order for offloading to MBS and HAPS-SMBS, which reduced energy consumption.
In another pertinent publication~\cite{HETS2023P}, the authors investigated scenarios where the entire
load was offloaded to the HAPS-SMBS instead of the MBS.
Moreover, \cite{Cihan} analyzed HAPS-SMBS deployment to accommodate unexpected traffic demands in TN, emphasizing capital and operational expenditures, while~\cite{9773096} explored how HAPS-SMBS could enhance sustainability in traditional micro and macro BS environments.

However, a limitation of previous studies is that they have not comprehensively considered both user association and energy optimization using standardized and realistic channel models. In particular, most of the earlier studies relied on simplified distance-based path-loss models, which are especially restrictive in HAPS-supported networks, where propagation characteristics differ significantly from terrestrial links. To address this gap, in the present study, we optimize energy consumption alongside user and cell association in vHetNets, enabling simultaneous offloading to HAPS and MBS. By explicitly adopting the 3GPP model for terrestrial and HAPS links~\cite{9443997},~\cite{9583591}, the proposed framework achieves realistic link-budget evaluations and accurate performance assessment. Using both TN and NTN resources, the framework enhances energy efficiency while maintaining outage-based quality of service (QoS), which constitutes the main motivation of this work.

\subsection{Contributions}
 
In this paper, we propose a novel HAPS-aided strategy for joint power consumption optimization and user association in CS. This innovative approach extends the concept of CS beyond traditional HetNets by integrating HAPS as an SMBS. HAPS brings the following unique advantages to the framework: 
(i) a footprint extending up to hundreds of kilometers in radius, far exceeding terrestrial macro coverage; 
(ii) persistent line-of-sight (LoS) connectivity, enabling more reliable and energy-efficient links; 
(iii) ability to operate on solar energy, reducing reliance on grid power and aligning with the emerging 3GPP “Networks Energy Savings” paradigm; and 
(iv) an additional high-capacity vertical tier for traffic offloading, which alleviates congestion in dense urban deployments and improves system scalability. 
This combination of wide-area coverage, reliable connectivity, energy efficiency, and offloading capacity distinguishes our framework from previous CS studies that were limited to terrestrial-only setups.
\textit{To the best of our knowledge, the present study is the first to jointly address power minimization and user association in a vHetNet, under standardized 3GPP channel models for both terrestrial~\cite{9443997} and HAPS links~\cite{9583591}, thereby ensuring realistic performance evaluation. This comprehensive perspective highlights the potential of HAPS to enhance energy efficiency and sustainability in future networks.}
The contributions of this paper are summarized as follows:

\begin{itemize} 
\item Using the potential of both NTN
and TN for efficient CS, we formulate an optimization problem to minimize network power consumption in a vHetNet that includes both HAPS and MBS, thereby achieving greater power conservation.

\item The optimization problem jointly considers the BS switching-off mechanism and user association to the best server, under the capacity constraints of the HAPS and MBS. We incorporate the following realistic channel models: the 3GPP TR 38.901 UMa model for terrestrial links~\cite{9443997} and the TR 38.811 LoS/NLoS model for HAPS~\cite{9583591}. This ensures that path-loss exponents, altitude, and LoS/NLoS variations are explicitly captured in the optimization framework, thus providing a realistic basis for performance evaluation.

\item We transform the formulated non-convex optimization problem into a mixed-integer programming (MIP) problem to derive an optimal solution. This approach not only minimizes power consumption, but also ensures that outage-based QoS requirements are met.

\item Through simulations, we demonstrate that the proposed method offers flexibility, enabling a range of solutions that balance power consumption and outage-based QoS. This flexibility allows network operators to prioritize either power consumption or user experience, providing solutions that can range, depending on network demands, from power-centric to QoS-centric.
\item We evaluate the performance of our proposed method under various network setups with differing degrees of heterogeneity. The results reveal that in contrast to the existing HAPS-aided CS algorithms typically limited to TN environments with only one type of SBS, our proposed algorithm is fully applicable to heterogeneous TN environments. Since it supports various types of SBSs, it is more adaptable and capable of managing the complex and diverse network architectures anticipated in 6G. 
\item The results of our extensive simulations demonstrate significant reductions in network power consumption, achieving up to 77\% savings compared to the All-ON method while maintaining high outage-based QoS standards for users.  Our algorithm achieves the lowest power consumption as compared to the four benchmark methods, providing thus an optimal balance between served traffic with outage-based QoS, energy efficiency, and computational complexity.

\end{itemize}

The remainder of this paper is organized as follows.
Section \ref{sec:system model} presents the system model, while Section \ref{sec:problem formulation} details the problem formulation. In Section \ref{sec:solution}, we introduce the HAPS-enhanced CS optimization algorithm. Performance evaluations are provided in Section \ref{sec:Performance-eva}. Finally, conclusions are drawn in Section \ref{sec:con}.

\section{System Model and Problem Constraints}\label{sec:system model}

\subsection{Network Model}
The network model investigated in this study is a vHetNet characterized by a high-density configuration that integrates ${s}$ SBSs with ${u}$ users within the coverage area of one MBS, along with one HAPS for downlink transmission. We also assume that ${N_A^k}$ is the number of antennas at MBS and HAPS, where $k \in \{M , H\}$ (i.e., ${k=M}$ for MBS and ${k=H}$ for HAPS). Another assumption is that all BSs connect to the users via an RF link. Each user can be served either by HAPS, MBS, or one of the SBSs. To enhance spectral efficiency and simultaneous user service, both MBS and HAPS employ Space Division Multiple Access (SDMA). This approach enables these stations to operate within the same frequency spectrum while using beamforming techniques. SBSs primarily provide data services and cater to user-specific requests, whereas the MBS and HAPS are responsible for maintaining consistent coverage, managing control signals, and providing data services.  
In addition, HAPS plays a central role in coordinating the offloading process by collecting load factors and link quality indicators from SBSs, together with the capacity status of the MBS. Acting as a centralized controller, it determines the ON/OFF states of SBSs and the corresponding offloading associations. While this introduces a coordination overhead, the exchanged information is lightweight and updated at a slow timescale or only when significant load variations occur, making it feasible given the substantial energy savings achieved. Consistent with emerging HAPS/NTN architectures, we assume that the HAPS obtains backhaul connectivity via high-capacity feeder links to a ground gateway, which provides a practical means to support the required control signaling. This overhead is necessary for efficient operation of the proposed cell-switching framework and is consistent with emerging architectures integrating NTN and TN, where HAPS serves a supervisory role. Importantly, although in our system model the HAPS footprint is restricted to match the MBS coverage area ($2\times2$ km) for fair comparison of offloading and energy-saving performance, in practice, HAPS can cover a footprint of up to $500$ km in radius~\cite{9583591}. This fundamental difference from traditional MBSs highlights scalability and unique potential of HAPS in future network deployments. Of note, HAPS does not simply replicate the role of the MBS in our framework. Rather, it acts as an \textit{SMBS}, using its persistent LoS connectivity to provide uniform coverage and support users in areas with poor or no TN coverage, while simultaneously offering substantial additional capacity to alleviate congestion in dense urban areas. This dual role differentiates HAPS from terrestrial MBSs and illustrates its unique value within the proposed cell-switching framework.

\begin{figure}[t]
\captionsetup{justification=raggedright, singlelinecheck=false}
\centerline{\includegraphics[width =8.cm ]{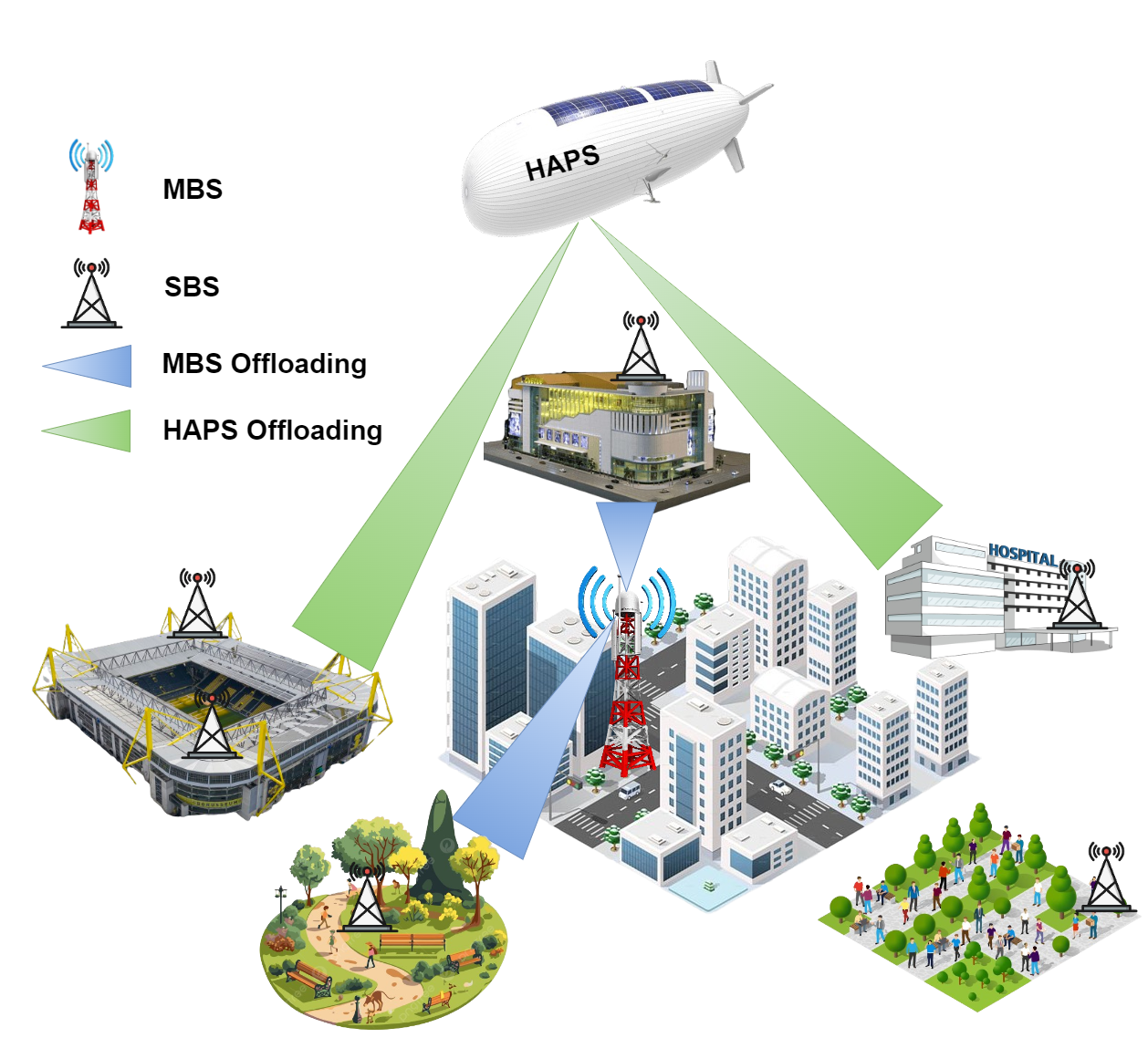}}
\caption{Illustration of a dynamic vHetNet configuration, highlighting CS opportunities between SBSs, MBS, and HAPS to optimize network performance and power efficiency.}
\label{Sys_model}
\end{figure}

Figure~\ref{Sys_model} provides a visual representation of the vHetNet configuration examined in this paper. The figure highlights the dynamic traffic offloading strategy, where HAPS plays a pivotal role in managing traffic for SBSs within its extensive coverage. Based on real-time assessments of traffic loads, channel conditions, and network power consumption, specific SBSs can redirect their users to the MBS or HAPS. This strategic offloading is designed to optimize the use of underutilized capacities to enhance the overall energy efficiency of the network. Offloading decisions are carefully made to ensure that the MBS and HAPS can handle additional traffic without compromising service quality. By dynamically adjusting user allocations among the MBS, HAPS, and active SBSs, this strategy minimizes total power consumption while maintaining high outage-based QoS under varying traffic conditions. This approach enables us to improve operational efficiency and achieve superior network performance with strict adherence to power efficiency criteria.

\subsection{RF Channel Model}

We adopt standardized 3GPP formulations for both terrestrial and HAPS links, capturing key effects such as altitude, LoS/NLoS probability, environment-specific path-loss exponents, and log-normal shadowing. This ensures that our evaluation is realistic and aligned with ongoing standardization efforts. In particular, the terrestrial channel follows the 3GPP Urban Macro (UMa) model~\cite{9443997}, while the HAPS channel is based on the LoS/NLoS formulation in 3GPP TR 38.811~\cite{9583591}.

\subsubsection{3GPP Channel Model for Terrestrial Platform}

Within the urban scenario outlined by the 3GPP standard model, the total path loss for the BS to user equipment (BS-UE) link can be expressed as follows~\cite{9443997}:
\begin{equation}
L = \rho ^{L} L^{L}  + \rho ^{N} L^{N},
\label{eq1}
\end{equation}

where ${\rho ^{L}}$ and ${\rho ^{N}}$ represent the line of sight (LoS) and non-line of sight (NLoS) probabilities, while ${L^{L}}$ and ${L^{N}}$ denote the associated losses in LoS and NLoS conditions, respectively.\\
The LoS probability in a terrestrial environment between the BS and the user can be given by~\cite{9443997}:
\begin{equation}
\rho ^{L}  = \left\{ {\begin{array}{*{20}c}
   {\begin{array}{*{20}c}
   1, \; \;\; \;\; \;\; \;\; \;\; \;\; \;\; \;\; \;\; \;\; \;\; \;\; \;\; \;\;\; & {d_{2D}  \le 18\,\mathrm{m},}  \\
\end{array}}  \\
   {\begin{array}{*{20}c}
   {\frac{{18}}{{d_{2D} }} + e^{( - \frac{{d_{2D} }}{{63}})(1 - \frac{{18}}{{d_{2D} }})} }, & {d_{2D}  \ge 18\,\mathrm{m},}  \\
\end{array}}  \\
\end{array}} \right.
\label{eq2}
\end{equation}
\begin{equation}
    \rho ^{N}  = 1 - \rho ^{L}, 
    \label{eq3}
\end{equation}
where ${d_{2D} }$ represents the two-dimensional (2D) distance between the BS and the user.
The path loss for LoS and NLoS links is provided as follows~\cite{9443997}:
\begin{equation}
L^{L}  = \left\{
\begin{array}{ll}
   L_{1}, &\;\; \;\;  10\,\mathrm{m} \le d_{2D} \le d_{b}, \\
   L_{2}, &\;\; \;\;  d_{b} \le d_{2D} \le 5\,\mathrm{km},
\end{array}
\right.
\label{eq4}
\end{equation}
where $L_{1}$ and $L_{2}$ represent the path loss expressions before and after the
breakpoint distance $d_b$, respectively. The breakpoint distance is defined as shown in \eqref{eq5}.
\begin{equation}
d_{b} = 4h'_{b} h'_{u} \frac{{f_c  \times 10^9 }}{C},
\label{eq5}
\end{equation}
where $f_c$ is the carrier frequency in GHz, and $C$ is the speed of light. 
The effective BS and UE antenna heights, denoted by $h'_{b}$ and $h'_{u}$, 
are defined as shown below.
\begin{equation}
h'_{b} = h_{b} - h_e, 
\label{eq6}
\end{equation}
\begin{equation}
h'_{u} = h_{u} - h_e,
\label{eq7}
\end{equation}
where $h_b$ and $h_u$ denote the BS and UE heights in meters, respectively, while $h_e$ is the effective environment height ($h_e=1  \mathrm{m}$ in urban settings). 

The detailed expressions for $L_{1}$ and $L_{2}$ are given as shown below.
\begin{equation}
L_{1}  = 28 + 22\log (d_{3D} ) + 20\log (f_c) + X,
\label{eq8}
\end{equation}
\begin{equation}
\begin{aligned}
\begin{split}
&L_{2}=  \\
&28 + 40\log (d_{3D} ) + 20\log (f_{c}) -9\log (d_{b}^2  + (h_{b}  - h_{u} )^2 )
+ X,
\end{split}
\end{aligned}
\label{eq9}
\end{equation}
\begin{equation}
L^{N}  = \max (L^{L} ,\hat L ^N ),
\label{eq10}
\end{equation}
\begin{equation}
\begin{aligned}
\begin{split} 
&\hat L ^N  =\\& 13.54 + 39.08\log (d_{3D} ) + 20\log (f_c) - 0.6(h_{u}  - 1.5) + X.
\end{split}
\end{aligned}
\label{eq11}
\end{equation}
Here, ${d_{3D}}$ is the three-dimensional (3D) distance between the BS and the user in meters. The shadow fading, represented by the log-normal random variable ${X}$, has a standard deviation of ${\sigma^{\text{L}} = 4 \text{ dB}}$ for LoS links and ${\sigma^{\text{N}} = 6 \text{ dB}}$ for NLoS links~\cite{9443997}.\\
Hence, the received power in dBm can be expressed as follows~\cite{9443997}:
\begin{equation}
    P_r=P_T+G_t+G_r-L,
    \label{eq12}
\end{equation}
where ${P_T}$ is the maximum transmit power at full capacity, ${G_t}$, and ${G_r}$ are the transmitter gain and receiver gains, respectively, and $L$ is calculated using \eqref{eq1}.

\subsubsection{3GPP Channel Model for HAPS Platform}

Considering the realistic
3GPP channel model, the LoS probability for the HAPS platform is presented based on the elevation angle of the HAPS relative to the terrestrial user and is denoted by ${\theta}$, as follows~\cite{9583591}:
\begin{equation}
\rho ^{L}  = a\theta ^b  + c,
\label{eq13}
\end{equation}
where $a$, $b$, and $c$ are environment-dependent parameters. For an urban area investigation, as reported in~\cite{9518388}, the values of $a$, $b$, and $c$ amount to 9.668, 0.547, and -10.58, respectively. The total path loss is calculated based on~\eqref{eq1}, with the path loss for LoS and NLoS links provided as follows~\cite{9583591}:
\begin{equation}
\begin{array}{*{20}c}
   {L^y}  = L_f + X^y  & {y \in \{ L,N\} },  \\
\end{array}
\label{eq14}
\end{equation}
\begin{equation}
L_f = 32.5 + 20\log (f) + 20\log (d_{3D} ).
\label{eq15}
\end{equation}
Here, $L_f$ represents the free-space path loss, and $X^y$ denotes the shadow fading, which follows a zero-mean normal distribution with a standard deviation of ${\sigma^{L} = 4 \text{ dB}}$ for LoS links and ${\sigma^{N} = 6 \text{ dB}}$ for NLoS links, specifically in urban areas~\cite{9518388}. 
The 3D distance between the HAPS and the terrestrial user is denoted by ${d_{3D}}$ and is given as follows~\cite{9583591}:
\begin{equation}
d_{3D}  = \sqrt {R_e^2 \sin ^2 (\theta ) + H_z^2  + 2H_z R_e }  - R_e \sin (\theta ),
\label{eq16}
\end{equation}
where ${R_e}$ and ${H_z}$ denote the Earth’s radius and HAPS altitude, respectively.\\
Therefore, the channel coefficient between the $n$-th antenna of the MBS or HAPS and the $i$-th offloaded user  from the $j$-th SBS, denoted by ${h_{i,j,n}^k }$, can be formulated as follows~\cite{9583591}:

\begin{equation}
\begin{array}{*{20}c}
   {h_{i,j,n}^k  = \frac{{f_{i,j,n}^k }}{{\sqrt {L_{i,j}^k} } }} & {k \in \{ M,H\} }. 
\end{array}
\label{eq17}
\end{equation}
Here, ${f_{i,j,n}^k}$ corresponds to the Rayleigh and Rician small-scale gain for ${k=M}$ and ${k=H}$, respectively. The path loss, ${L_{i,j}^k}$, represents the path loss of the $i$-th offloaded user from the $j$-th SBS to the BS of type $k$ and can be calculated for each user using~\eqref{eq1}, which incorporates log-normal shadowing.

\subsection{Network Power Consumption}
Based on the energy-aware radio and network technologies (EARTH) power consumption model~\cite{earth}, the power consumption by the $j$-th SBS at any given time, denoted by $ {P_j} $, is given by~\cite{7925662}:
\begin{equation}
P_j  = \left\{ {\begin{array}{*{20}c}
   {\begin{array}{*{20}c}
   {P_{O,j}  + \eta _j \lambda _j P_{T,j} }, & {0 < \lambda _j  < 1,}  
\end{array}}  \\
   {\begin{array}{*{20}c}
   {P_{S,j} },  \; \;   \; \; \; \; \; \; \;\; \; \; \; \; \; \; \; \; \; \; \; \; \; & {\lambda _j  = 0},  \\
\end{array}}  \\
\end{array}} \right.
\label{eq18}
\end{equation}
where $ {P}_{O,j} $ corresponds to the operational circuit power consumption, $ {\eta _j} $ is the power amplifier (PA) efficiency, $ {\lambda _j} $ is the load factor at the $j$-th SBS, $ {P}_{T,j} $ is the transmit power, and $ {P}_{S,j} $ is the sleep circuit power consumption. 
The network's instantaneous power consumption, denoted by $ {P} $, is calculated as follows:
\begin{equation}
P = P_M  + P_H  + \sum\limits_{j = 1}^s {P_j }, 
\label{eq19}
\end{equation}
where $ {P_M} $ and $ {P_H} $ are the MBS and HAPS power consumption at a given moment, respectively, and $ {s} $ is the number of SBSs in the deployed network.
Of note, $P_M$ and $P_H$ also follow the model described in~\eqref{eq18}, but do not have a sleep state, as they are always operational. 

\section{Problem Formulation}\label{sec:problem formulation}
We aim to identify the optimal state, balancing the activation and deactivation of SBSs, to minimize the overall network power consumption while adhering to users' outage-based QoS requirements. To this end, we define a state vector $\boldsymbol{\delta} = [\delta_1, \delta_2, \dots, \delta_s]$, representing the states of individual SBSs at a given time ${t}$. Each ${\rm \delta _j \in \{0,1}\}$ in the vector indicates the current state of a specific SBS, where 0 denotes OFF, and 1 denotes ON. Of note, the MBS and HAPS are constantly ON, signified by ${\delta _M = \delta _H = 1}$.

When ${\rm \delta _j}$ for a specific SBS changes from 1 to 0 at time $ {t}$, its associated users should be redirected to either the MBS or HAPS. This association is determined by evaluating the user channels with both the MBS and HAPS, calculating the received power from each, and selecting the BS with the maximum received power. 
The combined received signal by the $i$-th user (which initially belongs to the $j$-th SBS) from both the MBS and HAPS after employing linear precoding can be expressed as follows:
\begin{equation}
y_{i,j} = \sum\limits_{k \in \{ M,H \}} \mathbf{h}_{i,j}^k{}^H \left(\sum\limits_{i=1}^{u_{M}+u_{H}} \mathbf{w}_{i,j}^k x_{i,j}^k\right) + \nu_{i,j}.
\label{eq20}
\end{equation}
Here, ${\mathbf{h}_{i,j}^k = \left[ {h_{i,j,1}^k ,h_{i,j,2}^k ,...,h_{i,j,n}^k ,...,h_{i,j,N_A^k }^k } \right]^T}$ represents the RF channel vector between the $k$-th transmitter and the $i$-th user, offloaded from the $j$-th SBS to the MBS or HAPS (${k=M}$ or ${k=H}$), where $h_{i,j,n}^k$ is related to path loss through~\eqref{eq17}. Furthermore, ${\mathbf{w}_{i,j}^k} \in \mathbb{C}^{N_A^k }$ is the beamforming vector associated with the information signal ${x_{i,j}^k}$, and ${\nu_{i,j}}$ is the additive white Gaussian complex noise with variance ${\frac{{\sigma ^2 }}{2}}$ on each of its real and imaginary components. Next, \(u_M\) and \(u_H\) denote the total number of users supported by the MBS and HAPS, respectively, providing a combined total of \(u_M + u_H\) users in the inner summation.

Since the user should be associated with either the MBS or HAPS, we can rewrite~\eqref{eq20} using a binary variable ${s_{i,j}}$ as follows:
\begin{equation}
\begin{array}{l}
\begin{split}
&y_{i,j} = \\
&s_{i,j} \mathbf{h}_{i,j}^M{}^H \left(\sum\limits_{i=1}^{u_M} \mathbf{w}_{i,j}^M x_{i,j}^M\right) + (1 - s_{i,j}) \mathbf{h}_{i,j}^H{}^H \left(\sum\limits_{i=1}^{u_H} \mathbf{w}_{i,j}^H x_{i,j}^H\right) + \nu_{i,j},
\end{split}
\end{array}
\label{eq21}
\end{equation}
where binary variable ${s_{i,j}}$ takes a value of 1 when the $i$-th offloaded user is associated with the MBS, and 0 when associated with the HAPS.
To integrate the channel in our model, we can rewrite~\eqref{eq21} for the $p$-th user as follows:
\begin{equation}
\begin{array}{l}
\begin{split}
&y_{p,j} = s_{p,j}\left[\mathbf{h}_{p,j}^M{}^H \left(\mathbf{w}_{p,j}^M x_{p,j}^M + \sum\limits_{i=1,i\ne p}^{u_M}  \mathbf{w}_{i,j}^M x_{i,j}^M\right) \right] + \\&(1 - s_{p,j})\left[\mathbf{h}_{p,j}^H{}^H \left(\mathbf{w}_{p,j}^H x_{p,j}^H + \sum\limits_{i=1, i\ne p}^{u_H}  \mathbf{w}_{i,j}^H x_{i,j}^H\right) \right]+ \nu_{p,j}.
\end{split}
\end{array}
\label{eq22}
\end{equation}
For a massive multiple-input multiple-output (MIMO) system, as with an increase of ${N_A^k}$, the L2-norms of correlated vectors grow proportionally to ${N_A^k}$, while the inner products of uncorrelated vectors, by assumption, grow at a lower rate~\cite{5595728}. For large ${N_A^k}$, only the products of identical quantities remain significant. 
Therefore, by assuming a very large number of antennas for the HAPS and MBS, we can asymptotically assume that the small-scale fading vectors are orthogonal~\cite{6798744}.
\begin{equation}
\frac{{\boldsymbol{f}_{i,j}^k{}^H \boldsymbol{f}_{p,j}^k }}{{N_A^k }} \approx \left\{ {\begin{array}{*{20}c}
   {\begin{array}{*{20}c}
   {0,} & {p \ne i}  \\
\end{array}}  \\
   {\begin{array}{*{20}c}
   {1,} & {p = i},  \\
\end{array}}  \\
\end{array}} \right.
\label{eq23}
\end{equation}
where ${\boldsymbol{f}_{i,j}^k = \left[ {f_{i,j,1}^k,...,f_{i,j,N_A^k }^k } \right]^T}$. To manage inter-tier interference between the MBS and HAPS, we assume both employ matched filter precoding. With a large number of antennas, channel hardening leads to asymptotically orthogonal fading vectors~\cite{emil}, effectively mitigating interference. We can also assume that information on the received power from the HAPS and MBS is acquired during pilot transmissions using orthogonal sequences, which effectively suppresses different types of interference to a great extent. Given this orthogonality assumption, we use matched filter precoding for user $p$. In this technique, precoding vector \( \mathbf{w}_{p,j} \) is set to \( \bm{f}_{p,j}^k / N_A^k \). Now, using~\eqref{eq17} and this matched filter relationship, the received signal \( y_{p,j} \) in~\eqref{eq22} can then be expressed as follows:
\begin{equation}
y_{p,j}  =\frac{s_{p,j} x_{p,j}^M}{\sqrt{L_{p,j}^M}}   +\frac {(1 - s_{p,j} )x_{p,j}^H} {\sqrt{L_{p,j}^H}}.
\label{eq24}
\end{equation}
With this model, the received signal power \( P_{p,j}^r \) at the receiver of user $p$ is given by:
\begin{equation}
P_{p,j}^r  = \frac{s_{i,j} P_{T,M}}{U_M L_{i,j}^M}   + \frac{(1 - s_{i,j} )P_{T,H}}{U_H L_{i,j}^H} , 
\label{eq25}
\end{equation}
where \( P_{T,M} \) and \( P_{T,H} \) represent the maximum transmit powers for the MBS and HAPS, respectively, as defined in~\eqref{eq18}. \( U_M \) and \( U_H \) denote the total number of users that can be supported by MBS and HAPS. The first term in the equation corresponds to the received power from MBS, while the second term pertains to the received power from HAPS. Accordingly, we define \( P_{p,j,M}^r = \frac{P_{T,M}}{U_M L_{p,j}^M} \) and \( P_{p,j,H}^r = \frac{P_{T,H}}{U_H L_{p,j}^H} \). 
In our framework, the maximum transmit powers of the HAPS and MBS are set equal to ensure a fair comparison. However, the received powers $P_{p,j,H}^r$ and $P_{p,j,M}^r$ remain comparable because the distinct link budget formulations in Section~II-B capture the propagation characteristics: specifically, the HAPS benefits from LoS conditions with a lower path-loss exponent and higher antenna gain, while terrestrial links tend to experience NLoS losses with higher path-loss exponents and a lower antenna gain. To ensure outage-based QoS, we impose a constraint that the received power from either MBS or HAPS for each user exceeds a minimum threshold, \( P_{\text{min}} \), as shown in~\eqref{eq26}
\begin{equation}
\delta _j  + (1 - \delta _j )(s_{i,j} P_{i,j,M}^r  + (1 - s_{i,j} )P_{i,j,H}^r ) \ge (1 - \delta _j )P_{\min }. 
\label{eq26}
\end{equation}
Here, if \( \delta_j = 1 \), the inequality is inherently satisfied, indicating that the user’s outage-based QoS requirements are met without additional constraints. However, if \( \delta_j = 0 \) and the SBS is deactivated, then either \( P_{p,j,M}^r \) or \( P_{p,j,H}^r \) must meet or exceed \( P_{\text{min}} \), depending on the user’s association with MBS (\( s_{p,j} = 1 \)) or HAPS (\( s_{p,j} = 0 \)), to ensure the fulfillment of users' outage-based QoS requirements. Since the channel is reciprocal, and both the MBS and HAPS are equipped with highly capable receivers, we did not consider the same constraint for the uplink, as reliable decoding is inherently ensured under these conditions.
Moreover, during the CS ON and OFF process, additional constraints should be imposed on the loads of the MBS and HAPS  to ensure QoS for their users. The load on both MBS and HAPS at each time step is formulated as follows:
\begin{equation}
\lambda _M  = \lambda _{M,0}  + \sum\limits_{j=1}^{s}{(1 - \delta _j )} \left({\sum\limits_{i=1}^{u_j}  } s_{i,j} \lambda _{i,j}\right) \phi _{j,M}, 
\label{eq27}
\end{equation}
\begin{equation}
\lambda _H  = \lambda _{H,0}  + \sum\limits_{j=1}^{s}{(1 - \delta _j )} \left({\sum\limits_{i=1}^{u_j}  } (1-s_{i,j}) \lambda _{i,j}\right) \phi _{j,H}, 
\label{eq28}
\end{equation}
\begin{table*}[th!]
\caption{NOTATIONS AND DEFINITIONS}
\centering
\resizebox{\textwidth}{!}{ 
\begin{tabular}{ll|ll}
\hline\hline
\textbf{Symbol} & \textbf{Description} & \textbf{Symbol} & \textbf{Description} \\
\hline\hline
$u_j$ & Number of UEs in SBS $j$ & $u_M,\,u_H$ & Number of UEs served by MBS/HAPS \\
$U_M,\,U_H$ & Max UEs supported by MBS/HAPS & $h_b,\,h_u,\,h_e$ & BS / UE / Effective environment heights \\
$\rho^L,\,\rho^N$ & LoS / NLoS probability & $L,\,L^L,\,L^N$ & Total, LoS, NLoS path loss \\
$X$ & Shadow fading (log-normal) & $\theta$ & HAPS elevation angle \\
$R_e,\,H_z$ & Earth radius, HAPS altitude & $L_f$ & Free-space path loss (HAPS) \\
$\lambda_j$ & Load factor of SBS $j$ & $\lambda_{i,j}$ & Load from UE $i$ in SBS $j$ \\
$T_j$ & Total served traffic by SBS $j$ &
$C_j$ & Total capacity of SBS $j$ \\
$\lambda_M,\,\lambda_H$ & Aggregate load at MBS/HAPS & $\lambda_{M,0},\,\lambda_{H,0}$ & Initial load at MBS/HAPS \\
$\alpha$ & Load intensity scaling factor & $\phi_{j,M},\,\phi_{j,H}$ & Capacity ratios ($C_j/C_M$, $C_j/C_H$) \\
$T_{\text{QoS}}$ & Total served traffic with outage-based QoS & $\delta_j$ & State of SBS $j$ (ON=1/OFF=0) \\
$s_{i,j},\,s_j$ & UE association / group association & $z_j$ & Linearization variable for $\delta_j s_j$ \\
$\mathbf{h}_{i,j}^k$ & Channel vector (UE $i$, SBS $j$ to $k$) & $h_{i,j,n}^k$ & Channel coefficient (antenna $n$)\\
$f_{i,j,n}^k$ & Small-scale fading (Rayleigh/Rician) & $\mathbf{w}_{i,j}^k$ & Precoding vector at $k$ \\
$N_A^k$ & Number of antennas at MBS/HAPS & $L_{i,j}^k$ & Path loss (UE $i$, SBS $j$ to $k$) \\
$P_{p,j}^r$ & Received power at UE $p$ from SBS $j$ & $\nu_{i,j}$ & AWGN noise at UE $i$ from SBS $j$ \\
$P_{O,j}$, $P_{O,M}$, $P_{O,H}$ & SBS / MBS / HAPS operational power&
$P_{T,j}$, $P_{T,M}$, $P_{T,H}$ & SBS / MBS / HAPS transmit power\\
$P_{S,j}$ & SBS sleep power&
$\eta _j$, $\eta _M$, $\eta _H$& SBS / MBS / HAPS power amplifier efficiency \\
\bottomrule
\end{tabular}}
\label{tab:notation}
\end{table*}
where \(\lambda_{M,0}\) and \(\lambda_{H,0}\) represent the initial loads on the MBS and HAPS from the preceding time step, respectively. Parameter \(\lambda_{i,j}\) denotes the load contribution from the $i$-th user associated with the $j$-th SBS.  Here, \(u_j\) represents the number of users associated with the $j$-th SBS, and the total number of users across all SBSs is given by \( \sum_{j=1}^{s} u_j = u \). Furthermore, \(\phi_{j,M}\) and \(\phi_{j,H}\) indicate the relative capacity of the 
$j$-th SBS with respect to the MBS and HAPS, respectively. Specifically, \(\phi_{j,M} = {C_j}/{C_M}\), where \(C_j\) is the total capacity of the $j$-th SBS, and \(C_M\) is the total capacity of the MBS. Similarly, \(\phi_{j,H} = {C_j}/{C_H}\), where \(C_H\) denotes the total capacity of the HAPS. Here, by capacity, we refer to the maximum traffic a BS can support, determined by its available bandwidth and spectral efficiency.
Accordingly, the optimization problem can be formulated as follows:
\begin{IEEEeqnarray*}{lcl}\label{eq:P1}
&\underset{{s_{i,j}},\mathbf{\boldsymbol{\delta}}}{\text{minimize}}\,\, & ~P(s_{i,j},\boldsymbol{\delta} ) \,  \IEEEyesnumber \IEEEyessubnumber* \label{eq:P1_Obj}\\
    &\text{s.t.} & {\lambda _M}{\le 1,} \label{eq:P1_const1}\\
    && {\lambda _H}{\le 1,} \label{eq:P1_const2}\\
    && {{\delta _j,s_{i,j} \in \{ 0,1\}, }}  {{\;\;j=1,2,...,s}}, \label{eq:P1_const3}\\
    && {\eqref{eq26}-\eqref{eq28}.} \label{eq:P1_const4} 
\end{IEEEeqnarray*}
More specifically, the total power consumption 
$P$ depends on both the operational state vector of SBSs ($\boldsymbol{\delta}$) and user offloading decisions ($s_{i,j}$). Based on~\eqref{eq18} and~\eqref{eq19}, we can express $P$ as follows:
\begin{equation}
\begin{aligned}
\begin{split}
P(s_{i,j},\boldsymbol{\delta})=&{P_{O,M}  + \eta _M \lambda _M } P_{T,M} +{P_{O,H}  + \eta _H \lambda _H } P_{T,H}+ \\
&\sum\limits_{j = 1}^s {(P_{O,j}  + \eta _j \lambda _j  P_{T,j} )\delta _j +P_{S,j}(1-\delta _j)}.
\end{split}
\end{aligned}
\label{eq30}
\end{equation}
In problem~\eqref{eq:P1}, constraints~\eqref{eq:P1_const1} and~\eqref{eq:P1_const2} enforce that the total traffic load at the MBS and HAPS does not exceed their maximum capacity ($C_M$ and $C_H$, respectively). The load at a BS is defined as the ratio between its total allocated traffic (denoted by $T_j$) and its maximum capacity, i.e., $\lambda_j = T_j / C_j$. Therefore, these constraints ensure that the allocated traffic never surpasses the service limits of each BS, thereby maintaining QoS for all associated users. Constraint~\eqref{eq:P1_const3} specifies that both state variable \(\delta_j\) and association parameter \(s_{i,j}\) must take binary values.

Given the binary nature of \(\delta_j\) and \(s_{i,j}\), along with the non-convex constraint~\eqref{eq:P1_const4}, problem~\eqref{eq:P1} is classified as a mixed-integer nonlinear program (MINLP), which presents significant challenges for finding a global solution.
To address this complexity, we transform the problem into a mixed-integer programming (MIP) problem by linearizing the non-convex constraints. To solve the resulting MIP, we use solving constraint integer programs (SCIP), a robust framework designed to effectively handle various types of optimization problems, including constraint integer programming, MIP, and MINLP. SCIP’s efficiency and capability in managing complex, large-scale MIP problems make it a suitable choice for obtaining reliable solutions in our study.
For ease of presentation, we provide a list of the expressions used in this article in Table~\ref{tab:notation}.

\section{Proposed Solution} \label{sec:solution}
In this section, we transform problem~\eqref{eq:P1} into a linear form with respect to the optimization variables. 
To simplify the model, we treat all users associated with each SBS as a single group for offloading purposes, collectively directing them to the MBS or HAPS. This grouping approach is practical, assuming that users within the coverage area of a single SBS are geographically proximate, typically within a 50-meter radius of the SBS. Consequently, while small-scale fading might slightly differ among users, large-scale fading impacts them uniformly due to their similar distances from the SBS. Therefore, we aggregate individual loads, 
$\lambda_{i,j}$, into a single total load value for the $j$-th SBS, represented by 
$\lambda_j=\sum\limits_{i=1}^{u_j} {\lambda_{i,j}}$.
By treating users as a collective group, we simplify the problem by using a unified offloading decision variable, $s_j$, for each SBS instead of individual user-level decisions $s_{i,j}$.
Channel conditions are estimated based on a user located at the center of the SBS’s coverage area.

The primary challenge to solving this optimization lies in the non-convex constraint~\eqref{eq:P1_const4} and the bilinear products of binary variables, which make the original formulation a mixed-integer nonlinear program (MINLP). Specifically, the non-linearity arises from the multiplication of the cell switching variable $\delta_j$ and the offloading decision $s_j$. To address this concern, we first expand constraints~\eqref{eq26}-~\eqref{eq28}, as shown in~\eqref{eq31}-\eqref{eq33}.
\begin{equation}
\begin{aligned}
\begin{split}
\delta _j  + s_{j} P_{j,M}^r  + (1 - s_{j} )P_{j,H}^r  &- \delta _j s_{j} P_{j,M}^r  - \delta _j P_{j,H}^r + \delta _j s_{j} P_{j,H}^r \\ &\ge (1 - \delta _j )P_{\min }, 
\end{split}
\end{aligned}
\label{eq31}
\end{equation}
\begin{equation}
\lambda _M  = \lambda _{M,0}  + \sum\limits_{j=1}^{s} { {(s_{j}\lambda _{j} \phi _{j,M} - \delta _j s_{j}\lambda _{j} \phi _{j,M})} }\,  
\label{eq32}
\end{equation}
\begin{equation}
\begin{aligned}
\begin{split}
\lambda _H = &\lambda _{H,0} + \\ &\sum\limits_{j=1}^{s} (\lambda _{j} \phi _{j,H} - \delta _j\lambda _{j} \phi _{j,H}-s_{i,j}\lambda _{j} \phi _{j,H}+\delta _js_{i,j}\lambda _{j} \phi _{j,H} ).
\end{split}
\end{aligned}
\label{eq33}
\end{equation}
To linearize the bilinear terms, we introduce an auxiliary binary variable $z_j$ to represent the product $\delta_j s_j$. This technique replaces the nonlinear product with a new variable subject to additional linear constraints. By substituting $z_j$ into~\eqref{eq31}-\eqref{eq33}, the problem structure becomes compatible with MIP solvers:
\begin{equation}
\begin{aligned}
\begin{split}
\delta _j  + s_{j} P_{j,M}^r  + (1 - s_{j} )P_{j,H}^r  &- z_j P_{j,M}^r  - \delta _j P_{j,H}^r + z_j P_{j,H}^r\\
&\ge (1 - \delta _j )P_{\min }, 
\end{split}
\end{aligned}
\label{eq34}
\end{equation}
\begin{equation}
\lambda _M  = \lambda _{M,0}  + \sum\limits_{j=1}^{s} { {\left( s_{j}\lambda _{j} \phi _{j,M} - z_j\lambda _{j} \phi _{j,M}\right)} },  
\label{eq35}
\end{equation}
\begin{equation}
\begin{aligned}
\lambda _H &= \lambda _{H,0} + \sum\limits_{j=1}^{s} \left( \lambda _{j} \phi _{j,H} - \delta _j\lambda _{j} \phi _{j,H} - s_{j}\lambda _{j} \phi _{j,H} + z_j\lambda _{j} \phi _{j,H} \right).
\end{aligned}
\label{eq36}
\end{equation}
To ensure that \(z_j\) accurately reflects product \(\delta_j s_{j}\), constraints \eqref{eq37}-\eqref{eq39} are imposed:
\begin{equation}
z_j \leq s_{j},  
\label{eq37}
\end{equation}
\begin{equation}
z_j \leq \delta_j,  
\label{eq38}
\end{equation}
\begin{equation}
z_j \geq s_{j} + \delta_j - 1. 
\label{eq39}
\end{equation}
These constraints ensure that $z_j$ exactly captures the logical product $\delta_j s_j$: it takes the value 1 only if both $\delta_j$ and $s_j$ are 1, and 0 otherwise. As a result, the nonlinear terms are eliminated, and the overall problem~\eqref{eq:P1} is reformulated as an MIP:
\begin{IEEEeqnarray*}{lcl}\label{eq:P2}
    &\underset{\boldsymbol{s}, \boldsymbol{z},\mathbf{\boldsymbol{\delta}}}{\text{minimize}}\,\, & ~P(\boldsymbol{s}, \boldsymbol{z},\mathbf{\boldsymbol{\delta}} ) \,  \IEEEyesnumber \IEEEyessubnumber* \label{eq:P2_Obj}\\
    &\text{s.t.} & {\lambda _M}{\le 1,} \label{eq:P2_const1}\\
    && {\lambda _H}{\le 1,} \label{eq:P2_const2}\\
    && {{\delta _j,s_{j},z_{j}  \in \{ 0,1\}, }}  {{\;\; j=1,2,...,s }}, \label{eq:P2_const3}\\
    && {\eqref{eq34}-\eqref{eq38},\eqref{eq39}} \label{eq:P2_const4} 
\end{IEEEeqnarray*}
where $\boldsymbol{s} = [s_1, s_2, \dots, s_s]$ and $\boldsymbol{z} = [z_1, z_2, \dots, z_s]$.
Once the MIP framework is established, we introduce an optimization algorithm designed to achieve the optimal solution for the problem outlined in~\eqref{eq:P2}. The pseudocode for CS with offloaded user association is detailed in Algorithm 1. This algorithm addresses the optimization objective~\eqref{eq:P2_Obj} at each time step.
The pseudocode in Algorithm 1 orchestrates the process of CS complemented by the strategic offloading of user associations. This algorithm addresses the optimization objectives stated in~\eqref{eq:P2_Obj} for each time interval. Each time interval involves calculating potential received powers, forming the corresponding MIP, and solving it with the SCIP solver. The outcomes govern whether each SBS should offload its load to either the HAPS or the MBS and determine the activation state of the SBSs based on the optimized values of $\delta_{j}^*$ and $s^*_{j}$.
This approach ensures that the CS decisions are dynamically adapted to the changing network conditions, aiming to optimize network energy efficiency and maintain outage-based QoS across all users.

Along with the proposed HAPS-enhanced CS algorithm, we consider a specific scenario, referred to as HAPS-enhanced CS NoQoS, where the outage-based QoS constraint in~\eqref{eq26} is relaxed. This scenario allows for more SBSs to be switched off, as the offloading process focuses purely on minimizing power consumption, without guaranteeing user QoS. This case will be explored in subsequent simulation results to illustrate the trade-offs between power consumption and outage-based QoS.
\begin{algorithm}[t]
\caption{HAPS-enhanced CS Optimization Algorithm}
\DontPrintSemicolon  
\SetAlgoLined  
\KwData{$P_{O,j}$, $P_{O,M}$, $P_{O,H}$, $P_{T,j}$, $P_{T,M}$, $P_{T,H}$, $P_{S,j}$, $\eta _j$, $\eta _M$, $\eta _H$, $\lambda _{j}$, $\lambda _{M}$, $\lambda _{H}$, $P_\mathrm{min}$ }
\KwResult{$\boldsymbol{\delta}^*$,  $\boldsymbol{s}^*$}

\For{all time intervals $t$}{

    \For{all SBSs with index $j$}{
        Calculate $P^r_{j,H}$, $P^r_{j,M}$\;
        
        Form the MIP optimization problem in \eqref{eq:P2};
        
        Solve the MIP using SCIP and find $\boldsymbol{\delta}^*$, $\boldsymbol{s}^*$\;
        
        Offload all SBSs with $\delta_{j}^* = 0$ to the HAPS or MBS based on $s^*_{j}$\;
        
        Turn on or keep other SBSs active\;
    }
}

\label{alg1}
\end{algorithm}
\section{Performance Evaluation}\label{sec:Performance-eva}
In this section, we evaluate the effectiveness of our proposed CS strategy. The simulation parameters are detailed in Table~\ref{table:nonl}. Traffic distribution across the network is modeled using a 2D Gaussian function, which allows for precise control over the center (mean) and spread (variance) of the traffic load, representing the geographic distribution of users. Unless otherwise specified in the figure legends, all simulations assume a Gaussian model for traffic distribution. Each SBS is assigned a portion of the traffic based on its location, while the initial traffic conditions for MBS and HAPS are assigned separately. This configuration enables us to simulate diverse traffic scenarios and effectively evaluate the performance of the proposed CS method under varying load conditions. User mobility is statistically represented in our framework by varying the Gaussian 
distribution parameters (mean and variance) over time. While shifts in the mean correspond to the 
movement of traffic hotspots, changes in the variance represent user dispersion or 
concentration. These variations directly affect the SBS load factors $\lambda_j$, 
and the proposed cell switching framework dynamically adapts to these changes, ensuring 
robust performance under mobility.
In the simulation setup, 49 SBSs are uniformly distributed across the area, with the MBS located at the center of the traffic load distribution (mean) and the HAPS positioned at an altitude of $20$~km directly above the network’s origin.
We selected $s=49$ to represent a sufficiently large and dense grid network that stresses the system and highlights the robustness and scalability of the proposed framework.
Figure~\ref{snapshot}, which illustrates the network layout, shows the SBSs, MBS, and HAPS positions.

\begin{table}[t] 
\centering 
\caption{SIMULATION SETUP}
\resizebox{.5\textwidth}{!}{
\begin{tabular}{lll}
\hline\hline 
Parameters & Notation & Typical Values\\
\hline\hline
Area size& D & $2000\; m\times 2000\; m$
\\
Carrier frequency& ${f_c}$ & 2.5 GHz 
\\
Number of SBSs& S & 49
\\
Small cell radius& R & 50 m
\\
Speed of light& ${C}$ &  299792458 m/s
\\
Transmit power of the MBS & ${P_T^M}$& 43 dBm~\cite{9583591}
\\
Transmit power of the HAPS  & ${P_T^H}$ & 43 dBm~\cite{9583591}
\\
Antenna gain of the MBS  & ${G_T^M}$ & 8 dBi~\cite{9443997}
\\
Antenna gain of the HAPS  & ${G_T^H}$ & 43.2 dBi~\cite{9583591}
\\
Shadow fading standard deviation (LoS)&$\sigma ^{L}$  & 4 dB ~\cite{9443997}
\\
Shadow fading standard deviation (NLoS)& $\sigma ^{N}$ &  6 dB~\cite{9443997}
\\
Receiver sensitivity& $P_{\min }$ & -70 dBm
\\
Capacity of the SBS   & ${C_j}$ & 5
\\
Capacity of the MBS  & ${C_M}$ & 20
\\
Capacity of the HAPS & ${C_H}$ & 40
\\

\hline
\end{tabular}}
\label{table:nonl}
\vspace{-.5cm}
\end{table}

\subsection{Benchmarking}
To evaluate the performance of the proposed HAPS-enhanced CS algorithm, the following three different benchmark methods are compared:

\subsubsection{All-ON}
In this method, no switching off is applied, and all SBSs remain active at all times. As a result, there are no concerns regarding QoS, since all users are served by the closest SBS with the strongest received power, determined based on~\eqref{eq12}. This method serves as a baseline for evaluating power consumption, as it represents the scenario with no power-saving strategies.

\subsubsection{Sorting}Following~\cite{6679195}, 
 in this method, SBSs are sorted based on their load factors, $\lambda$. The SBSs with the lowest load are then switched off sequentially until the capacity of the MBS is fully used, leaving the remaining SBSs active. Given the power consumption profiles presented in~\eqref{eq18}, along with the characteristics of different types of BSs listed in Table \ref{table: power profile}, the MBS generally consumes more power than SBSs for equivalent traffic loads. Therefore, the algorithm prioritizes switching off SBSs with smaller loads, aiming to minimize overall energy consumption. However, this method does not consider the QoS for users offloaded from the switched-off SBSs, which could lead to poor signal reception from the MBS for users located farther from it.
 
\subsubsection{Terrestrial CS }As reported in~\cite{Metin_VFA_CellSwitch}, this baseline method optimizes the network’s power consumption by switching off SBSs and offloading traffic solely to the MBS, without the involvement of HAPS. Unlike the optimization problem defined in~\eqref{eq:P1}, the Terrestrial CS approach does not consider the 
received power constraint~\eqref{eq26}, nor does it require binary variable ${s_j}$
for HAPS offloading decisions. Consequently, the only constraints are on the MBS load factor, as described in~\eqref{eq:P1_const1}, and on the SBS state, as described in~\eqref{eq:P1_const3} (${\delta _j \in \{ 0,1\}}$), resulting in an MIP problem.

\subsubsection{Exhaustive Search (ES)} This comprehensive approach explores all possible ON/OFF configurations for the SBSs to determine the optimal state vector. The objective is to identify the configuration that would minimize power consumption. However, computational complexity of this method is extremely high, as it requires evaluating ${2^{s}}$ possible configurations, where ${s=49}$ SBSs are considered in our setup (see Table~\ref{table:nonl}). Given this vast search space, implementing ES is not feasible for practical scenarios. Therefore, while it serves as a theoretical optimal benchmark, we compared its time complexity with our proposed methods.

\subsection{Performance Metrics}
In this section, we present the key metrics used to evaluate the performance of the proposed solution and the benchmark methods.

\subsubsection{Power Consumption}\label{subsubsec:pow-con}
Total power consumption, calculated using~\eqref{eq30}, serves as an important metric for evaluating the performance of the network. This metric is particularly relevant to evaluate sustainability of the proposed solution, as reducing power consumption directly correlates with a more environmentally friendly and cost-effective network operation.

\subsubsection{Total Served Traffic with QoS}\label{subsub:QoS}
The total served traffic with outage-based QoS, $T_{QoS}$, is defined as the cumulative traffic handled by all BSs while maintaining the required outage-based QoS. This is calculated using \eqref{eq41}.
\begin{equation}
T_{QoS}  = T_M  + T_H  + \sum\limits_{j = 1}^s {T_j }, 
\label{eq41}
\end{equation}
where $T_M$ and $T_H$ denote the traffic served by the MBS and HAPS, respectively, while $T_j$ represents the traffic served by the $j$-th SBS. The traffic served with outage-based QoS by each SBS depends on its operational state (ON or OFF) and the received power of the $i$-th offloaded users if the $j$-th SBS is OFF ($P_{i,j,k}^r,\;\;  k\in {M, H}$). Specifically, the values are defined as follows:
\begin{equation}
   T_j  = 
\begin{cases}
        {C_j \lambda _j }, & {\delta _j  = 1},\\
        {C_j \lambda _j }, & {\delta _j  = 0 \;\; and \;\; P_{i,j,k}^r  \ge P_{\min } },\\
        0, & {\delta _j  = 0\;\; and \;\;P_{i,j,k}^r <  P_{\min } }.
    \end{cases}
    \label{eq42}
\end{equation}
This formulation assumes that, if an SBS is active ($\delta _j  = 1$), it fully serves its assigned load, $C_j \lambda _j$. However, if the SBS is switched OFF ($\delta _j  = 0$), and the received power at the user falls below the threshold $P_{\min }$ (i.e., $P_{i,j,k}^r <  P_{\min }$), the served traffic for that SBS is set to zero, as an outage occurs due to insufficient signal quality. This metric thus evaluates how effectively the network serves traffic while maintaining outage-based QoS, particularly under various SBS configurations and offloading scenarios.

\subsubsection{Energy Efficiency}\label{subsub
} Energy efficiency is defined as the ratio of total served traffic with QoS to the total power consumption. It reflects how well the network converts power into served traffic while maintaining QoS (see \eqref{eq43}).
\begin{equation} 
\textit{Energy Efficiency} = \frac{T_{QoS}}{{P}}. \label{eq43} \end{equation} 
This metric helps to evaluate the trade-off between energy savings and service delivery.

\subsubsection{Time Complexity}
Time complexity, or computational complexity, measures the execution time of the algorithm, which is critical for assessing its scalability as the number of SBSs increases. This metric highlights how network expansion affects the algorithm's performance, particularly with regard to how quickly it can respond to changes in traffic demand and offloading scenarios. With an increase in the network size, this metric becomes essential for ensuring that the algorithm remains efficient and practical for large-scale deployments.

\begin{table}[t] 
\caption{DIFFERENT BSs POWER PROFILE}
\centering  
\resizebox{.5\textwidth}{!}{
\begin{tabular}{lllll} 
 \hline\hline 
& & &Power Consumption& \\
BS Type&Efficiency&Transmit&Operational&Sleep
\\ 
&$\eta $&$P_{T}$$[W]$&$P_{O}$$[W]$&$P_{S}$$[W]$
\\  
\hline\hline
HAPS&4.7&20&130&75
\\
Macro~\cite{earth}&4.7&20&130&75
\\
RRH~\cite{earth}&2.8&20&84&56
\\
Micro~\cite{earth}&2.6&6.3&56&39 
\\
Pico~\cite{earth}&4&0.13&6.8&4.3
\\
Femto~\cite{earth}&8&0.05&4.8&2.9
\\[1ex] 
    \hline 
    \end{tabular}}
    \label{table: power profile} 
    \vspace{-0.5cm}
    \end{table}

\subsection{Case Studies}
The developed benchmark methods and the proposed HAPS-enhanced CS algorithm are evaluated under two distinct case studies, referred to thereafter as Case Study A and Case Study B. These case studies are designed to compare the performance of the methods under different network configurations.

\subsubsection{Case Study A} This is a straightforward setup where only micro-type SBSs are deployed in the network. In addition, the sleep mode power consumption of the SBSs is assumed to be zero, meaning that inactive SBSs do not contribute to the total network power consumption~\cite{Metin_VFA_CellSwitch}. This setup allows for a simpler evaluation of the algorithms, focusing on basic energy savings and performance optimization.

\subsubsection{Case Study B} 

This case study represents a more complex scenario that involves a network configuration with multiple types of SBSs: micro, remote radio head (RRH), pico, and femto. These SBSs are deployed across the network with nearly equal distribution, consisting of 13 micros and 12 of each other type.
Unlike Case Study A, Case Study B incorporates a more realistic energy consumption model, where the sleep mode power consumption is based on the values in Table~\ref{table: power profile}, rather than assuming zero power. In addition, the different SBS types have varying power consumption profiles, capacities, and coverage areas. Accordingly, the proposed HAPS-enhanced CS algorithm is tested in a more complex scenario that requires sophisticated offloading and CS decisions, accounting for dynamic interactions in a diverse HetNet environment.

\begin{figure*}[t]
    \centering
    \captionsetup{justification=raggedright, singlelinecheck=false}
    \subfloat[HAPS-enhanced CS NoQoS.]{%
        \includegraphics[width=.33\textwidth]{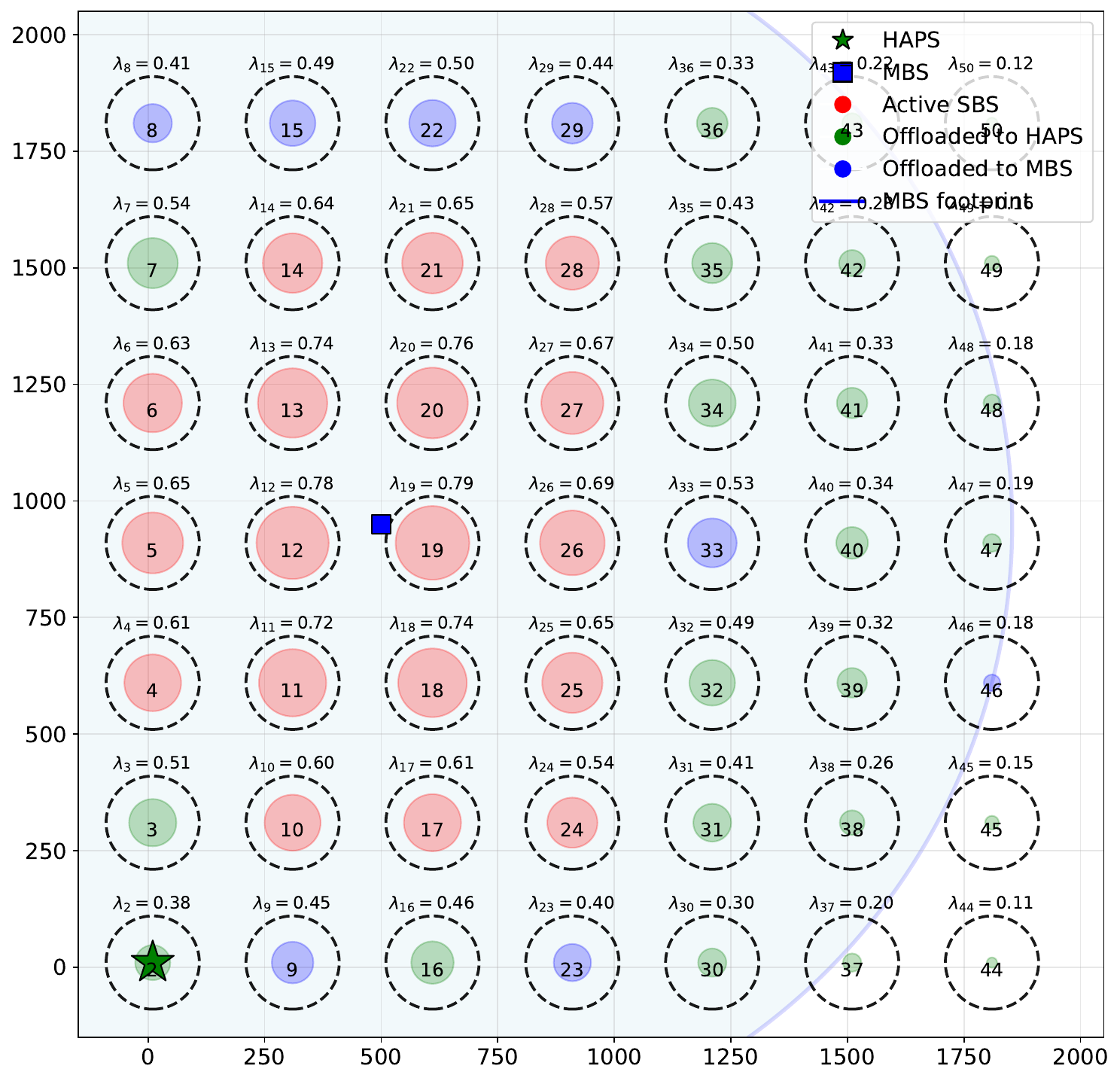}%
        \label{snapshot:a}}%
    \subfloat[HAPS-enhanced CS, $P_\mathrm{min}=-80 \;\mathrm{dBm}$.]{%
        \includegraphics[width=.33\textwidth]{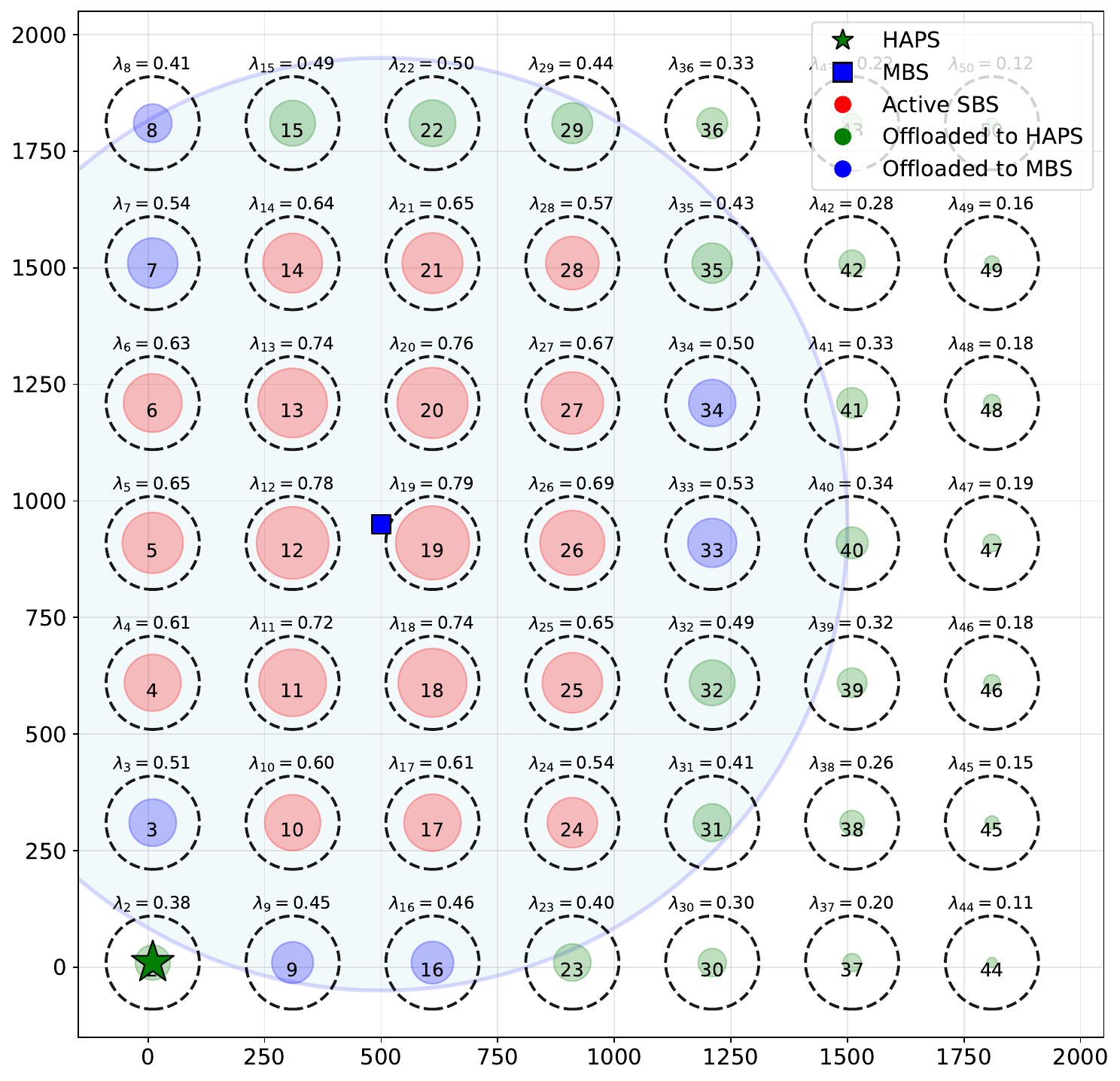}%
        \label{snapshot:b}}%
    \subfloat[HAPS-enhanced CS, $P_\mathrm{min}=-60 \;\mathrm{dBm}$.]{%
        \includegraphics[width=.33\textwidth]{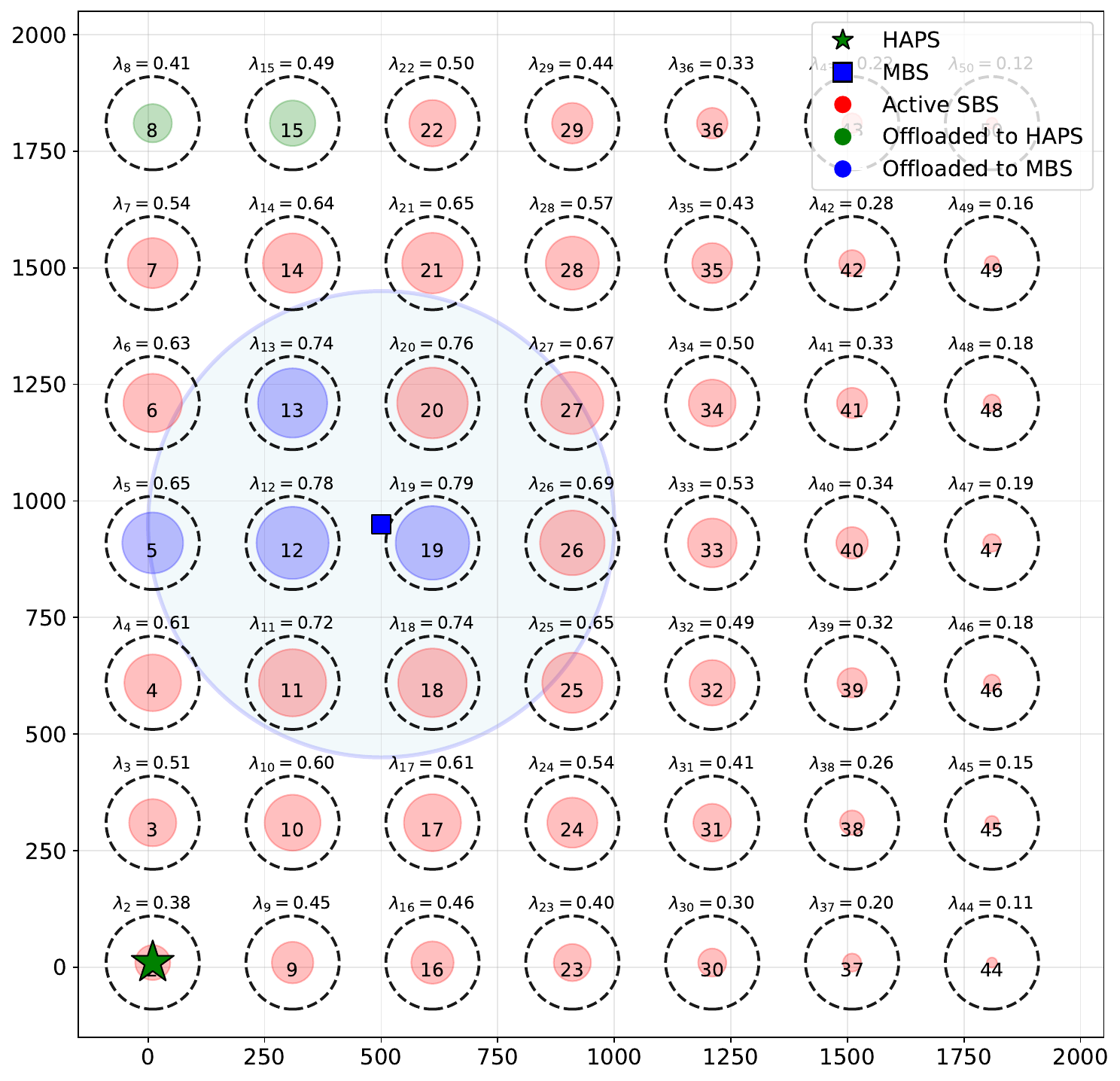}%
        \label{snapshot:c}}%
    \caption{Snapshots of the network status for HAPS-enhanced CS NoQoS and HAPS-enhanced CS under different $P_\mathrm{min}$ values (Case Study A). Outer dotted circles represent the coverage areas of SBSs, while inner solid circles denote their load ($\lambda$ values). The color of each SBS indicates its status: active, offloaded to the MBS, or offloaded to the HAPS. The large blue circle illustrates the SBSs offloaded to the MBS under the corresponding outage-based QoS condition.}
        \label{snapshot}
\end{figure*}

\subsection{Simulation Results}
This section reports the simulation results to evaluate the performance of the proposed HAPS-enhanced CS algorithm using the defined metrics and compares the results against the benchmark methods.

To provide a clearer understanding of the proposed approaches, we present snapshots of the network status for the HAPS-enhanced CS NoQoS and HAPS-enhanced CS under various $P_\mathrm{min}$ values in Fig.~\ref{snapshot}. In these snapshots, the outer dotted circles represent the coverage area of SBSs, while the radius of inner solid circles is related to their load, shown by $\lambda$ values. The gaps between SBS footprints are covered by the MBS and the HAPS. Importantly, the coverage area of the HAPS is much larger than the $2$~km grid considered here and is assumed to encompass the entire area, while the offloading of SBSs to the HAPS depends on the QoS enforcement.
The color of these circles corresponds to the SBS status: active, offloaded to the MBS, or offloaded to the HAPS, as indicated in the legend.
In Fig.~\ref{snapshot:a}, the HAPS-enhanced CS NoQoS approach, which operates independently of 
$P_\mathrm{min}$, deactivates as many SBSs as the capacities of the HAPS and MBS allow, focusing solely on minimizing power consumption without considering QoS constraints. Consequently, this approach deactivates 31 SBSs.
By contrast, the remaining subfigures illustrate the HAPS-enhanced CS approach under varying 
$P_\mathrm{min}$ values. For the low $P_\mathrm{min}$ setting ($-80\;\mathrm{dBm}$), similarly to HAPS-enhanced CS NoQoS, $31$ SBSs are deactivated due to the minimal QoS restrictions. At the higher considered $P_\mathrm{min}$ value ($-60\;\mathrm{dBm}$), only $6$ SBSs are deactivated, as the algorithm prioritizes maintaining user QoS over power savings.
The large blue circle in each subfigure indicates the area from which several SBSs are offloaded to the MBS under the corresponding QoS condition.
Due to a higher path loss exponent and NLoS conditions, MBS channels experience a significant path loss, making $P_\mathrm{min}$ adjustments more impactful on the MBS. Conversely, HAPS benefits from LoS channels that maintain lower path loss exponents, making HAPS coverage less sensitive to changes in $P_\mathrm{min}$.

\begin{figure}[t]
    \centering
    \captionsetup{justification=raggedright, singlelinecheck=false}
    \subfloat[Case Study A: homogeneous network with micro SBSs.]{%
        \includegraphics[width=.4\textwidth]{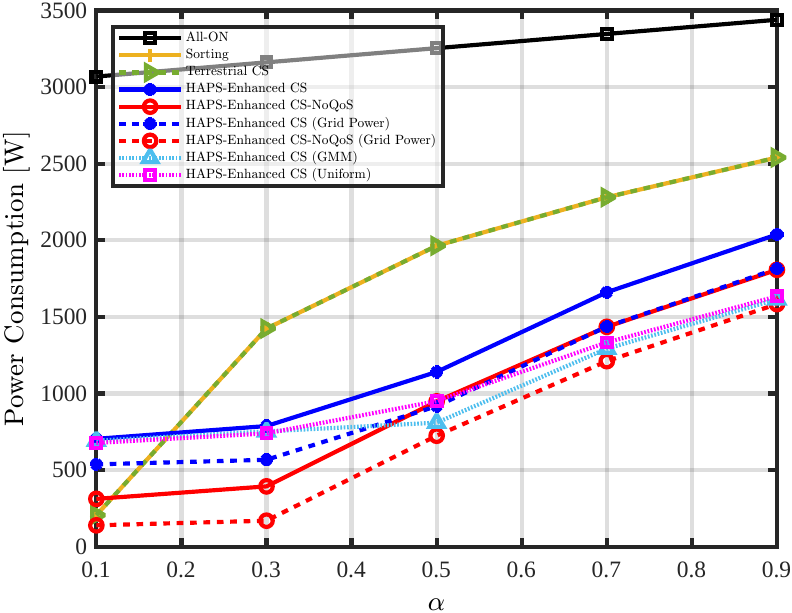}%
        \label{Pvsload:a}
    }\\
       \subfloat[Case Study B: heterogeneous network with micro, pico, and femto SBSs.]{%
        \includegraphics[width=.4\textwidth]{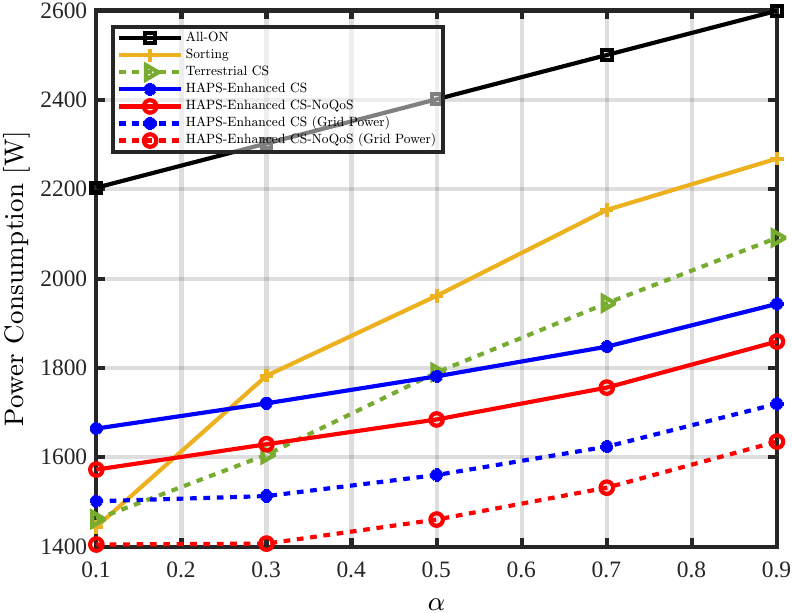}%
        \label{Pvsload:b}}
    \caption{Total power consumption vs. load intensity for various CS methods, with $P_\mathrm{min}=-70\; \mathrm{dBm}$.}
   \label{Pvsload}
\end{figure}

Figure~\ref{Pvsload} shows the total power consumption of the network as a function of the load intensity, denoted by $\alpha$.
Load intensity represents the scaling factor for the peak user demand in the Gaussian traffic distribution, allowing us to compare the performance of the proposed HAPS-enhanced CS and HAPS-enhanced CS NoQoS algorithms with that of benchmark methods.
Figure~\ref{Pvsload:a} shows the results for Case Study A, where all SBSs are of the micro type. 
In line with our expectation, Fig.~\ref{Pvsload:a} shows that the All-ON method has the highest power consumption across all load intensities, since it keeps all SBSs active regardless of traffic demand. This baseline clearly demonstrates the inefficiency of the continuous activation of all SBSs. The performance gap between the All-ON method and the other methods is the largest at lower load intensities, which is consistent with the EARTH power model, indicating that lightly loaded BSs are less energy efficient compared to the heavily loaded ones.
With an increase in the load intensity $\alpha$, there are fewer opportunities to switch off SBSs, resulting in a reduced performance gap. For instance, at $\alpha=0.1$, the proposed HAPS-enhanced CS algorithm reduces power consumption by $77\%$ as compared to the All-ON method, whereas at $\alpha=0.9$, this reduction drops to $40\%$. These reduction percentages were computed as the relative difference between the total power consumption of the All-ON baseline and that of the proposed HAPS-enhanced CS method, normalized to the baseline value.
The Sorting and Terrestrial CS methods show overlapping results, initially outperforming the proposed HAPS-enhanced CS algorithm. This is so because, at lower load intensities, the terrestrial offloading mechanism effectively manages traffic. Introducing the HAPS on this level leads to additional power consumption due to its power amplifier profile. Importantly, both HAPS and MBS have higher power amplifier efficiencies (${\eta _M}$ and ${\eta _H}$, as indicated in Table~\ref{table: power profile}) as compared to SBSs (${\eta _j}$), which increases energy usage when offloading to these higher-capacity nodes.
However, as the load intensity increases (around $\alpha=0.2$), the proposed HAPS-enhanced CS outperforms the Terrestrial CS due to the HAPS's greater capacity and coverage, enabling more effective offloading and reducing overall power consumption. The performance gap peaks at around $\alpha=0.5$, where the HAPS-enhanced CS achieves the highest power savings as compared to Terrestrial CS. As the load intensity continues to rise, the performance difference narrows, similarly to the trend observed with the All-ON method.
The HAPS-enhanced CS NoQoS, another variant of the proposed algorithm, achieves an even lower power consumption as it does not impose QoS constraints on offloaded users, allowing more SBSs to be switched off. For comparison, relative to the All-ON method, the HAPS-enhanced CS NoQoS reduces total power consumption by approximately $90\%$ at $\alpha=0.1$ and by $47\%$ at $\alpha=0.9$, reflecting its stronger energy-saving potential.
Furthermore, grid power consumption, excluding the HAPS’s solar power usage, is calculated for both HAPS-enhanced approaches, demonstrating the HAPS’s essential role in reducing grid power reliance.
In addition to the Gaussian distribution, a two-component Gaussian Mixture Model (GMM) and a uniform distribution are also evaluated in Fig.~\ref{Pvsload:a} to assess the robustness of the proposed HAPS-enhanced CS algorithm. The results show that even under more complex or evenly distributed user scenarios, the proposed method maintains strong performance across all load intensities.

The results shown in Fig.~\ref{Pvsload:b}, which depicts the total power consumption for Case Study B, follow a trend similar to those shown in Fig.~\ref{Pvsload:a}. However, there is a notable increase in power consumption across all approaches (except for the All-ON method that exhibits lower power consumption due to the inclusion of pico and femto SBSs that have a lower power profile as compared to micro SBSs) due to the non-zero sleep mode power consumption (see Table~\ref{table: power profile}).
In contrast to Case Study A, the Sorting method's performance in Case Study B deviates from the optimized Terrestrial CS approach. This discrepancy arises from greater heterogeneity of the network in Case Study B, where different SBS types have distinct power profiles, capacities, and coverage areas. As a result, sorting SBSs based solely on their load becomes less effective in minimizing power consumption, given the diverse power demands of different SBS types.

The HAPS-enhanced CS algorithm continues to demonstrate a better performance in this scenario, not only because of its optimization framework, but also because the HAPS’s high-altitude LoS coverage reduces the dependency of QoS on SBS location within the $2$ km grid. This makes QoS degradation less sensitive to whether near or far SBSs are switched off, unlike MBS offloading where distance has a much stronger effect. 
Meanwhile, the HAPS-enhanced CS NoQoS approach still achieves the lowest power consumption, focusing exclusively on minimizing power by switching off more SBSs, without considering the QoS constraint. 
Overall, these results highlight the essential role of HAPS in enabling significant energy savings while maintaining user QoS. By contrast, in the absence of HAPS,  the QoS trade-off under CS becomes significantly harsher, as terrestrial links alone are more prone to high path loss and coverage limitations.

\begin{figure}[t]
\centerline{\includegraphics[width=7.3cm, height=5.5cm]{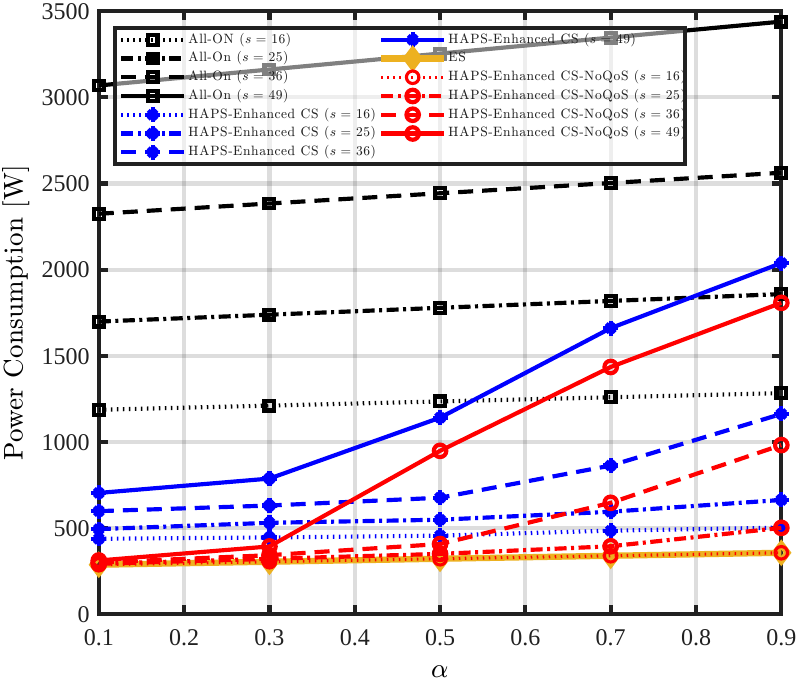}}
\caption{Impact of the number of SBSs ($s=16,25,36,49$) on power consumption.}
\label{Diff_SBSs}
\end{figure}

To further investigate the impact of SBS density, in Fig.~\ref{Diff_SBSs}, we compare networks with different numbers of SBSs ($s=16,25,36,49$). In line with our expectation, a higher number of SBSs results in a greater overall power consumption, since more BSs are active in the All-ON baseline. More importantly, the relative energy savings of the proposed HAPS-enhanced CS approaches are greater when fewer SBSs are deployed. With lower SBS densities, the MBS and HAPS have more available capacity to offload traffic, leading to flatter and more linear growth in power consumption with increasing load. By contrast, with higher SBS densities, offloading flexibility is reduced, and the slope of power consumption increases more sharply with the load intensity. For instance, at $\alpha = 0.9$, the relative power saving decreases from about $72\%$ for $s=16$ to about $40\%$ for $s=49$.
In addition, for the case of $s=16$, we include the ES results as a performance benchmark. As ES explores all $2^{s}$ SBS state combinations, it yields the optimal solution in terms of power consumption, but becomes computationally infeasible for larger networks (see Fig.~10, where the execution time for $s=16$ is already $1 \mathrm{s}$). The results show that the HAPS-enhanced CS NoQoS approach achieves the same optimal power consumption as ES across all load intensities, demonstrating that our method can achieve optimal power savings while remaining computationally efficient and scalable to larger deployments.

\begin{figure}[t]
    \centering
    \captionsetup{justification=raggedright, singlelinecheck=false}
    \subfloat[Total served traffic with QoS for Case Study A.]{%
        \includegraphics[width=7.1cm, height=5.6cm]{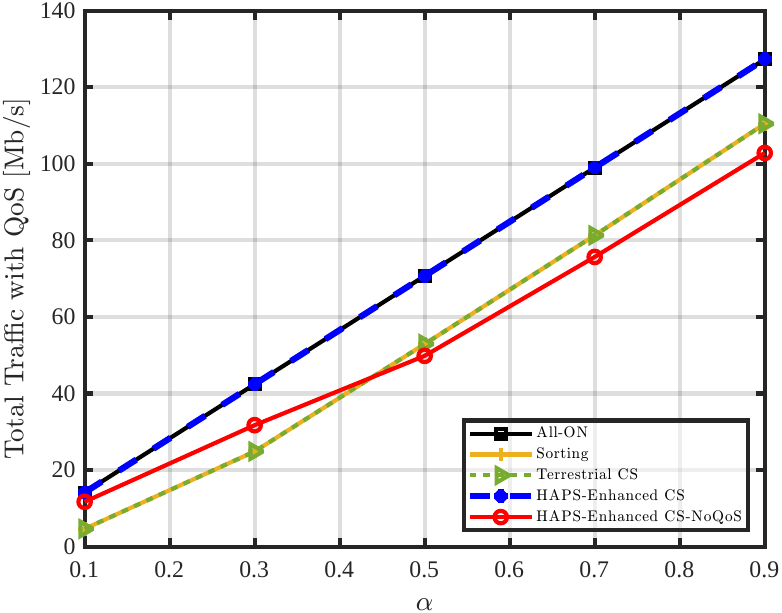}%
        \label{Tvsload:a}}%
        \\
    \subfloat[Total served traffic with QoS for Case Study B.]{%
        \includegraphics[width=7.1cm, height=5.6cm]{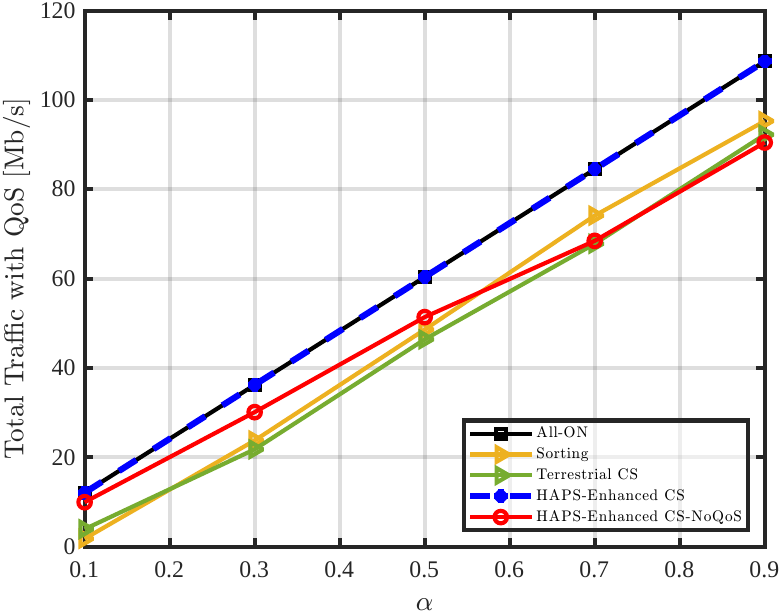}%
        \label{Tvsload:b}}%
    \caption{Total served traffic with QoS vs. load intensity for various CS methods, with $P_\mathrm{min}=-70\; \mathrm{dBm}$.}
    \label{Tvsload}
\end{figure}

Figure~\ref{Tvsload} shows the total served traffic with QoS, as defined in Section \ref{subsub:QoS}, across different load intensities. Figure~\ref{Tvsload:a} shows the results for Case Study A, while Fig.~\ref{Tvsload:b} those of Case Study B. 
In Case Study A (Fig.~\ref{Tvsload:a}), the proposed HAPS-enhanced CS approach maintains total served traffic with QoS close to that of the All-ON method, achieving the maximum possible value. This result highlights the primary advantage of the HAPS-enhanced CS approach: namely, that it significantly reduces network power consumption while ensuring QoS for all offloaded users. By contrast, the Terrestrial CS and HAPS-enhanced CS NoQoS approaches fall short of this maximum value, as neither enforces the QoS constraint defined in~\eqref{eq26}. 
At lower load intensities, the HAPS-enhanced CS NoQoS approach outperforms Terrestrial CS. This can be attributed to the fact that, in Terrestrial CS, offloading is limited to the MBS. Many distant and lightly loaded SBSs offload their users to the MBS, leading to QoS violations due to poor channel conditions for these far-off users. To compare, the HAPS-enhanced CS NoQoS approach offloads most users to the HAPS, benefiting from consistent LoS links that offer relatively uniform QoS for all users.
With an increase in load intensity, the performance dynamics change. Specifically, Terrestrial CS begins to outperform HAPS-enhanced CS NoQoS due to MBS capacity constraints, which limit the number of offloaded SBSs. This constraint helps to preserve the QoS for users served by their original SBSs. By contrast, HAPS-enhanced CS NoQoS continues to aggressively offload users to maximize power savings, thereby compromising QoS.
For Case Study B (Fig.~\ref{Tvsload:b}), the overall trend is similar to Case Study A, and our proposed method achieves the highest traffic. However, the transition point, where Terrestrial CS starts outperforming HAPS-enhanced CS NoQoS, shifts from $\alpha=0.45$  to approximately $\alpha=0.8$. This indicates that HAPS-enhanced CS NoQoS sustains higher total served traffic with QoS over a broader range of load conditions, even within the more complex HetNet environment of Case Study B.

\begin{figure}[h]
    \centering
    \captionsetup{justification=raggedright, singlelinecheck=false}
    \subfloat[Total power consumption for Case Study A.]{%
        \includegraphics[width=.4\textwidth]{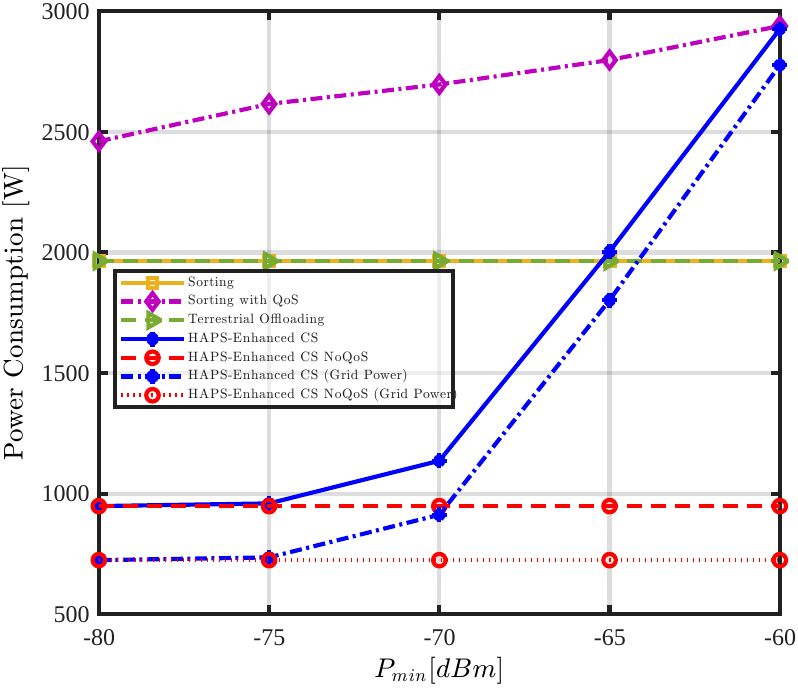}%
        \label{Pvspmin:a}}%
        \\
    \subfloat[Total power consumption for Case Study B.]{%
        \includegraphics[width=.4\textwidth]{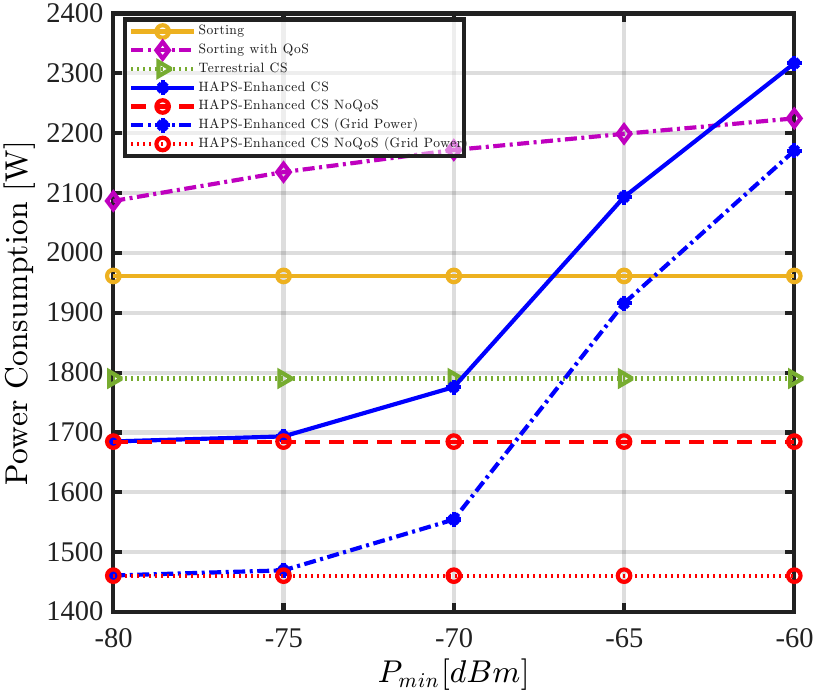}%
        \label{Pvspmin:b}}%
    \caption{Total power consumption vs. $P_\mathrm{min}$
for the proposed HAPS-enhanced CS methods and benchmark algorithms, with 
$\alpha=0.5$.}
    \label{Pvspmin}
\end{figure}

Figure~\ref{Pvspmin} illustrates the total power consumption of the network as a function of $P_\mathrm{min}$, with simulations conducted for a fixed load intensity ($\alpha=0.5$). In addition to the previously described baselines, we also include a QoS-aware variant of the Sorting heuristic, in which an SBS is switched off only if its offloaded users still satisfy the outage-based QoS requirement $P_\mathrm{min}$. As expected, this QoS-aware Sorting scheme exhibits higher power consumption than the original Sorting method because some lightly loaded SBSs can no longer be switched off due to the $P_\mathrm{min}$ constraint.
As can be seen in Fig.~\ref{Pvspmin:a}, the results for Case Study A show that the schemes whose power consumption varies with $P_\mathrm{min}$ are the HAPS-enhanced CS and the QoS-aware Sorting heuristic, both of which explicitly enforce the outage-based QoS constraint. By contrast, the original Sorting, Terrestrial CS, and HAPS-enhanced CS NoQoS are independent of $P_\mathrm{min}$, resulting in constant power consumption regardless of the actual $P_\mathrm{min}$ value. As also seen in Fig.~\ref{Pvsload:a} for Case Study A, the results for Sorting and Terrestrial CS overlap. The QoS-aware Sorting variant exhibits higher power consumption than the original Sorting method, since some lightly loaded SBSs can no longer be switched off when their offloaded users would violate the $P_\mathrm{min}$ constraint.
Initially, at lower power thresholds ($P_\mathrm{min} \le -75\;\mathrm{dBm}$), the HAPS-enhanced CS approach behaves similarly to HAPS-enhanced CS NoQoS because the $P_\mathrm{min}$ condition is not stringent and most users have a received power exceeding this threshold. As $P_\mathrm{min}$ becomes stricter (i.e., increases), the differences between the two approaches become more noticeable. While the HAPS-enhanced CS reduces the number of switched-off SBSs to ensure outage-based QoS, leading to increased power consumption, the HAPS-enhanced CS NoQoS continues to prioritize minimizing power consumption, switching off as many SBSs as possible without considering QoS, as long as the offloading capacities of MBS and HAPS are not exceeded. This illustrates the trade-off between power savings and QoS management in the proposed methods. By choosing an appropriate value for $P_\mathrm{min}$, the HAPS-enhanced CS approach can strike a balance between these two factors. Importantly, these simulation results also demonstrate that HAPS-enhanced CS is a general method, with $P_\mathrm{min}$ serving as an additional degree of freedom to manage both QoS and power consumption. If QoS is not a concern, setting $P_\mathrm{min}$ to a very low value effectively transforms HAPS-enhanced CS into the HAPS-enhanced CS NoQoS approach.

Figure~\ref{Pvspmin:b} shows the results for Case Study B. As in the results shown in Fig.~\ref{Pvsload:b}, Sorting and Terrestrial CS methods no longer overlap due to the heterogeneous SBS power profiles, which makes the simple Sorting heuristic less efficient than the optimized Terrestrial CS. The QoS-aware Sorting variant further increases the total power consumption, since additional SBSs must remain active when their users cannot be offloaded without violating the $P_\mathrm{min}$ constraint. Nevertheless, the HAPS-enhanced CS approach continues to provide the most favorable trade-off between power savings and outage-based QoS, demonstrating its adaptability in complex network conditions.

\begin{figure}[t]
    \centering
    \captionsetup{justification=raggedright, singlelinecheck=false}
    \subfloat[Total served traffic with QoS for Case Study A.]{%
        \includegraphics[width=.4\textwidth]{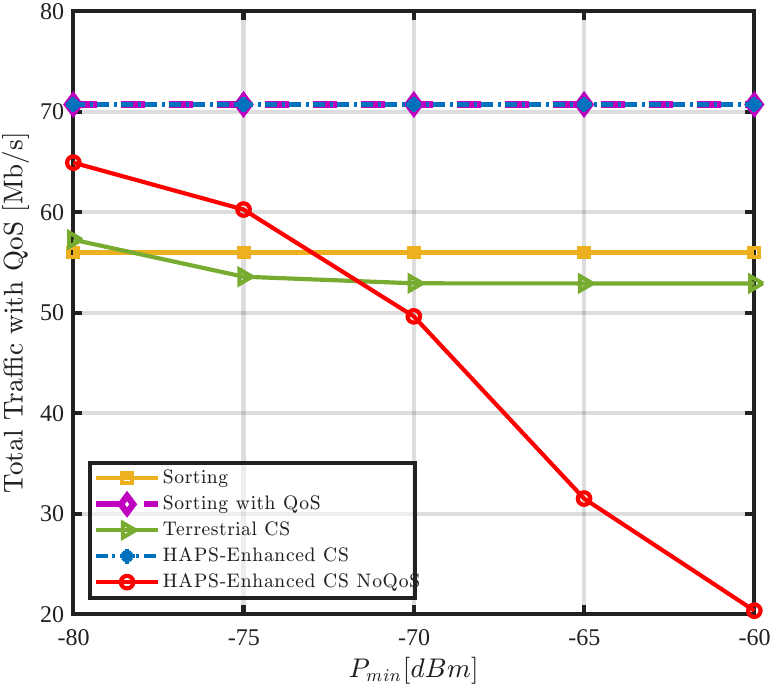}%
        \label{Tvspmin:a}}%
        \\
    \subfloat[Total served traffic with QoS for Case Study B.]{%
        \includegraphics[width=.4\textwidth]{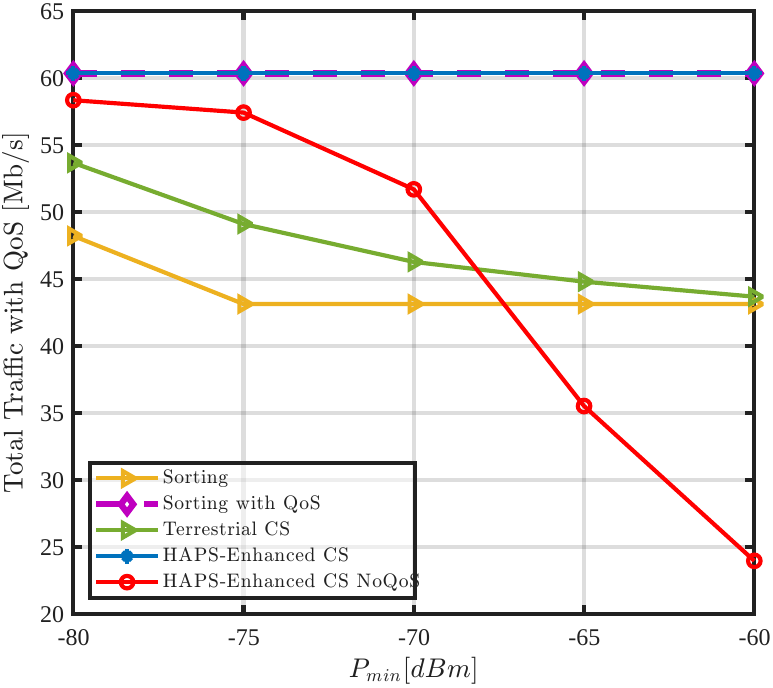}%
        \label{Tvspmin:b}}%
    \caption{Total served traffic with QoS as a function of different $P_\mathrm{min}$ for various methods, with 
$\alpha=0.5$.}
    \label{Tvspmin}
\end{figure}

Figure~\ref{Tvspmin} shows the total served traffic with QoS as a function of different $P_\mathrm{min}$ values for a fixed load intensity ($\alpha=0.5$). Figure~\ref{Tvspmin:a} presents the results for Case Study A, while Fig.~\ref{Tvspmin:b} depicts the results for Case Study B. 
As seen in both figures, the HAPS-enhanced CS approach maintains QoS across all 
$P_\mathrm{min}$ values, demonstrating robustness in ensuring user satisfaction even as QoS criteria become stricter. This results in a decreasing number of switched-off SBSs, reflecting the adaptability of the approach to stricter QoS requirements and highlighting the trade-off between power efficiency and QoS management.
Conversely, the HAPS-enhanced CS NoQoS approach prioritizes minimizing the total power consumption and remains unaffected in the CS process by the changes in $P_\mathrm{min}$. As such, the total served traffic with QoS decreases with an increase of $P_\mathrm{min}$, emphasizing its sole focus on power reduction.
For the Terrestrial CS approach, the number of switched-off SBSs remains constant, as it does not account for received power constraints. In Case Study A (Fig.~\ref{Tvspmin:a}), the served traffic with QoS remains relatively stable across 
$P_\mathrm{min}$ values, since offloading is limited to the MBS. The Sorting method closely tracks Terrestrial CS here, showing slightly higher stability across most $P_\mathrm{min}$ ranges.
In Case Study B (Fig.~\ref{Tvspmin:b}), as $P_\mathrm{min}$ becomes more stringent, the served traffic of Terrestrial CS declines more noticeably, while Sorting degrades even further. This occurs because ranking SBSs solely by load poorly interacts with heterogeneous SBS power profiles, leading to offloading choices that more frequently fail the QoS threshold. This further underscores the improved performance of the proposed HAPS-enhanced CS in complex environments.
In both case studies, the QoS-aware Sorting variant achieves the highest served traffic with outage-based QoS, as the $P_\mathrm{min}$ constraint prevents switching off SBSs whose users would violate the threshold.

\begin{figure}[t]
\centerline{\includegraphics[width=7.5cm, height=6.3cm]{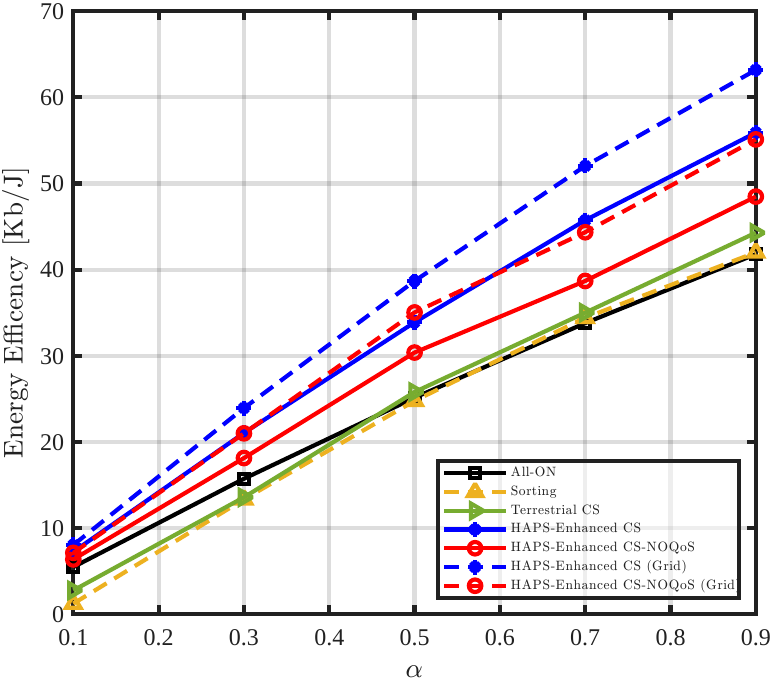}}
\caption{Energy efficiency vs. load intensity for Case Study B, with $P_\mathrm{min}=-70\; \mathrm{dBm}$.}
\label{EEvsload}
\vspace{-.5cm}
\end{figure}
Figure~\ref{EEvsload} shows energy efficiency as a function of the load intensity for Case Study B. As the load intensity increases, energy efficiency also rises across all approaches. This increase is driven by a more substantial growth in total served traffic as compared to power consumption. The HAPS-enhanced CS approach consistently achieves the highest energy efficiency, particularly at higher load intensities, due to its effective offloading strategy. The HAPS-enhanced CS NoQoS also shows good performance, but is slightly lower than the HAPS-enhanced CS. Conversely, the All-ON method demonstrates the lowest energy efficiency, as it keeps all SBSs active regardless of user demand. The Terrestrial CS and Sorting approaches also perform poorly, ranking close to the All-ON method, which is reasonable given their traffic-with-QoS and power-consumption characteristics.

\begin{figure}[t]
\centerline{\includegraphics[width=7.5cm, height=6.3cm]{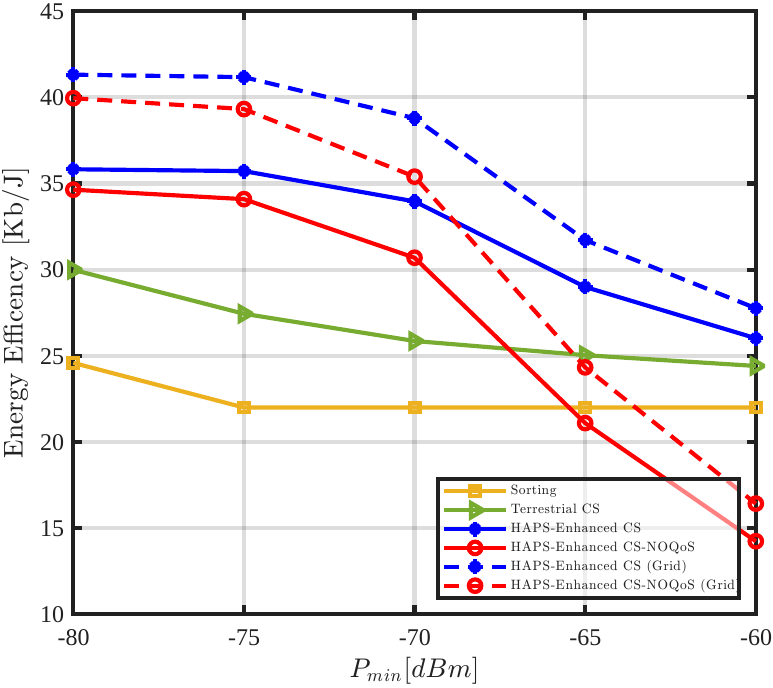}}
\caption{Energy efficiency vs. $P_\mathrm{min}$ for Case Study B, with $\alpha=0.5$.}
\label{EEvspmin}
\end{figure}

Figure~\ref{EEvspmin} shows the energy efficiency of different approaches as a function of $P_\mathrm{min}$ for Case Study B. 
The HAPS-enhanced CS method consistently achieves the highest energy efficiency across all 
$P_\mathrm{min}$ values, thus outperforming the HAPS-enhanced CS NoQoS variant. This is so because the increase in served traffic with QoS outweighs the rise in power consumption, making HAPS-enhanced CS more effective in maintaining energy efficiency, especially as QoS requirements become stricter.
For higher $P_\mathrm{min}$ values, even the Terrestrial CS approach starts to outperform the HAPS-enhanced CS NoQoS, highlighting the impact of stricter QoS requirements on energy efficiency.
For the Sorting method, since its power consumption remains constant and does not depend on $P_\mathrm{min}$, its energy efficiency trend simply follows the same decline as its served traffic with QoS (Fig.~\ref{Tvspmin:b}). Furthermore, when considering grid power consumption alone (excluding the HAPS solar power), both the HAPS-enhanced CS and HAPS-enhanced CS NoQoS methods achieve higher energy efficiency, as expected.

\begin{figure}[t]
\centerline{\includegraphics[width=7.5cm, height=6.3cm]{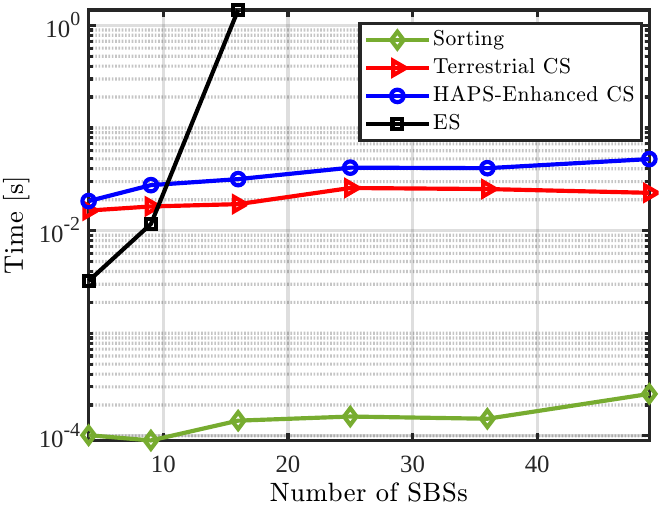}}
\caption{Time complexity: Execution time vs. number of SBSs.}
\label{Time_complexit}
\vspace{-.5cm}
\end{figure}
Figure~\ref{Time_complexit} compares the time complexity of different CS algorithms as a function of the number of SBSs, with time measured in seconds on a logarithmic scale. The figure presents the results of the following four approaches: Sorting, Terrestrial CS, HAPS-enhanced CS, and ES.
In line with our expectation, ES exhibits the highest computational complexity, with time growing exponentially as the number of SBSs increases. However, for smaller SBS counts, ES initially shows lower complexity than optimization-based methods like HAPS-enhanced CS and Terrestrial CS. This outcome can be attributed to the fact that, at lower SBS numbers, the optimization overhead outweighs the cost of enumerating all possible configurations in ES. As the SBS count grows, the exponential nature of ES makes it impractical for larger networks.
By contrast, the HAPS-enhanced CS approach demonstrates a more gradual increase in complexity, scaling approximately linearly with the number of SBSs. Although it starts with a higher initial computational time as compared to ES, its complexity remains manageable, making it suitable for larger networks.
Similarly, Terrestrial CS maintains linear complexity, since it focuses solely on offloading to the MBS. This makes this approach well-suited for large-scale networks, as the computational time only increases moderately with a rising number of SBSs.
The Sorting method is the most computationally efficient, maintaining minimal time complexity regardless of the SBS count. It achieves this by sorting SBSs based on load and sequentially switching them off. However, this approach is less effective for HetNets, as it does not account for varying power consumption profiles across different SBS types, which limits its effectiveness in such scenarios.
The computational energy of the solver is negligible (average runtime $\approx 0.05 \mathrm{s}$ for $s>16$ SBSs), which is orders of magnitude smaller than the operational power of SBSs ($>200$ W, see Fig.~\ref{Diff_SBSs}) and is therefore excluded from the overall power consumption results.

\section{Conclusion}\label{sec:con}
 In this study, we proposed a HAPS-enhanced CS algorithm to optimize energy consumption and user association in vHetNets. By integrating the HAPS with the MBS and diverse SBSs, our approach demonstrated significant power savings up to 77\% while consistently maintaining high levels of user outage-based QoS.
Our method effectively balances energy efficiency and outage-based QoS by dynamically switching off underutilized SBSs based on real-time traffic loads, channel conditions, and network capacity constraints. Unlike the algorithms proposed in previous studies, our algorithm comprehensively addresses both user association and energy optimization within heterogeneous TN, showcasing adaptability across different SBS types and complex vHetNet architectures.
Future research could focus on integrating additional NTN elements to further enhance comprehensive CS capabilities, improving scalability and performance under diverse and dynamic network conditions.

\section*{Acknowledgment}

This study is supported in part by the Connected and Autonomous Vehicles (TrustCAV) CREATE Program funded by the Natural Sciences and Engineering Research Council of Canada (NSERC).

\ifCLASSOPTIONcaptionsoff
  \newpage
\fi
\small
\input{IEEE-Jornal.bbl}

\end{document}